\documentclass{article}
\usepackage{cite,graphicx,amsmath,amssymb,epsfig}

\def\beq{\begin{equation}}
\def\eeq{\end{equation}}

\def\beqn{\begin{eqnarray}}
\def\eeqn{\end{eqnarray}}
\newcommand{\mbf}[1]{\mbox{\boldmath $#1$}}
\newcommand{\bk}{\mbf{k}}
\newcommand{\ba}{\mbf{a}}
\newcommand{\bb}{\mbf{b}}

\newcommand{\bq}{\mbf{q}}

\newcommand{\bp}{\mbf{p}}

\begin{document}

\title{\Large \bf $N=4$ supersymmetric Yang Mills 
scattering amplitudes\\ at high energies: the Regge cut contribution}
\author{\large J.~Bartels$^{1}$, L.~N. Lipatov$^{1,2}$, A.~Sabio~Vera$^{3}$ 
\bigskip \\
{\it  $^1$~II. Institut Theoretical Physics, Hamburg University, Germany} \\
{\it  $^2$~St. Petersburg Nuclear Physics Institute, Russia}\\
{\it  $^3$~CERN, Geneva, Switzerland, \&}\\
{\it Instituto de F{\' i}sica Te{\' o}rica UAM/CSIC, Universidad}\\ 
{\it Aut{\' o}noma de Madrid, E-28049 Madrid, Spain}}

\maketitle

\vspace{-9cm}
\begin{flushright}
{\small CERN--PH--TH/2008--125}\\
{\small DESY--08--073}\\
{\small IFT--UAM/CSIC--09--11}
\end{flushright}

\vspace{7cm}
\begin{abstract}
\noindent
We further investigate, in the planar limit of $N=4$ supersymmetric Yang Mills 
theories, 
the high energy Regge behavior of six-point MHV scattering amplitudes. 
In particular, for the new Regge cut contribution found in our previous 
paper, we compute in the leading logarithmic approximation (LLA) the energy 
spectrum of the BFKL equation in the color 
octet channel, and we calculate explicitly the two loop corrections to 
the discontinuities of the amplitudes for the
transitions $2 \to 4$ and $3 \to 3$. We find an explicit solution
of the BFKL equation for the octet channel for arbitrary momentum transfers
and investigate the intercepts of the Regge singularities in this channel.
As an important result we find 
that the universal collinear and infrared singularities of the BDS 
formula are not affected by this Regge-cut contribution. Any improvement of 
the BDS formula should reproduce this cut to all orders in the coupling.

\end{abstract}

\section{Introduction}

In a recent work~\cite{Bartels:2008ce} we have investigated the high energy 
Regge behavior of MHV scattering amplitudes in the planar limit of  
$N=4$ supersymmetric Yang Mills Theories, and we have found that,
for n-point amplitudes with $n>5$ beyond the one loop approximation, the 
simple factorizing structure of the Bern-Dixon-Smirnov (BDS) 
conjecture~\cite{Bern:2005iz}
is not valid. In detail, it was shown that for the cases of the transitions
$2\rightarrow 4$ and $3\rightarrow 3$ their 
factorized form is violated by Regge cut contributions which satisfy the BFKL 
equation~\cite{BFKL} in the color octet channel. These terms are obtained from 
specific single energy discontinuities, and in the scattering amplitudes 
they become visible in particular physical kinematic regions only.   
In the one loop approximation, these terms are correctly contained  
in the BDS formula, but in higher orders they cannot be cast into the 
simple exponential form conjectured by Bern et al.   
 
In this paper we further investigate these Regge cut contributions.
We study the BFKL equation in the color octet state, and we compute 
the two-loop expressions for the $2\to4$ and $3\to3$ amplitudes.
In particular, we show that the collinear and infrared divergences 
of the BDS formula are not affected by the Regge cut contributions.

The paper is organized as follows. In section 2 we briefly review the 
derivation of the factorization-breaking contributions, and we 
write down the expression for the Regge-cut contribution, using the 
calculus of complex momenta. Sections 3 - 5 are devoted to the 
detailed investigation of this Regge-cut contribution: we first (section 3) 
study the structure of the infrared singularities, we then (section 4) 
compute the two loop expressions for the cut contributions, and finally 
in section 5 we obtain  
the explicit solution of the BFKL equation for the octet channel. 
In the final section we present 
conclusions and further strategies. 
Solutions of the BFKL equation for the forward case are presented in an 
appendix.        
  
\section{The Regge cut contribution: review and representation in terms of 
complex momenta} 

\begin{figure}[ht]
\centerline{\epsfig{file=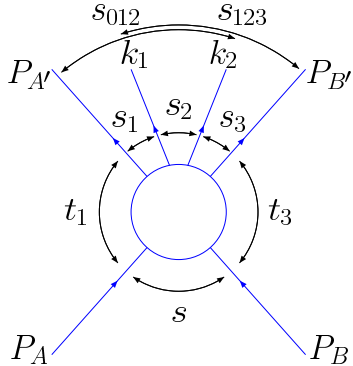,width=5cm,
angle=0,bbllx=0,bblly=506,bburx=237,
        bbury=738,clip=}}
\caption{Kinematics of the $2 \to 4$ amplitude} 
\label{24kinematics}
\end{figure}

\begin{figure}
\centerline{\epsfig{file=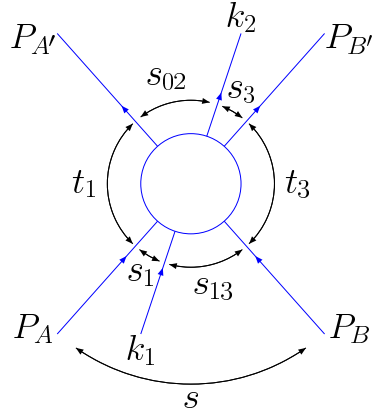,width=5cm,
angle=0,bbllx=130,bblly=460,bburx=355,
        bbury=706,clip=}}
\caption{Kinematics of the $3 \to 3$ amplitude} 
\label{33kinematics}
\end{figure}

\begin{figure}
\centerline{\epsfig{file=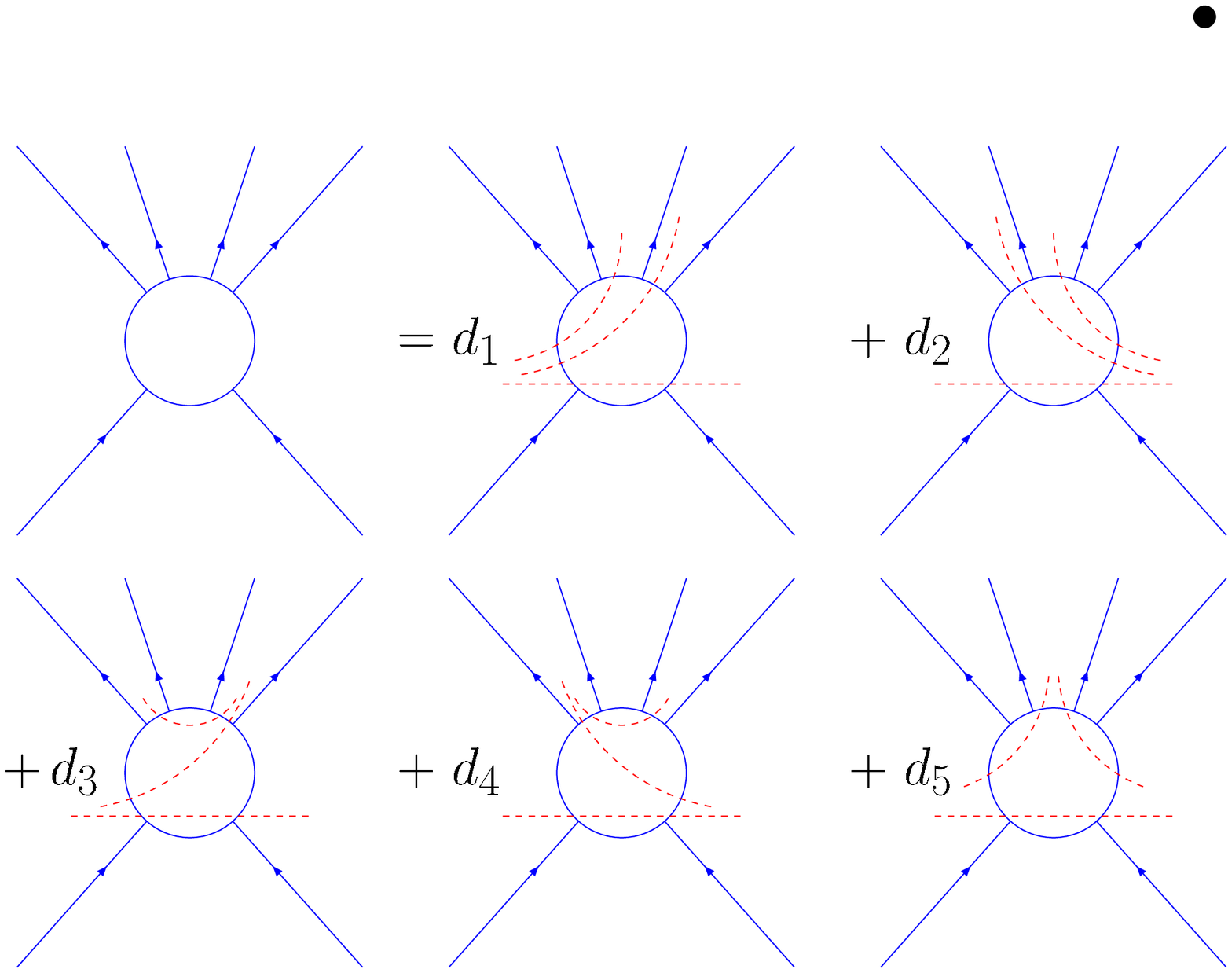,width=10cm,angle=0,bbllx=0,bblly=300,bburx=584,
        bbury=700,clip=}}
\caption{Analytic representation of the amplitude $M_{2\rightarrow 4}$} 
\label{disper24}
\end{figure}

\begin{figure}
\centerline{\epsfig{file=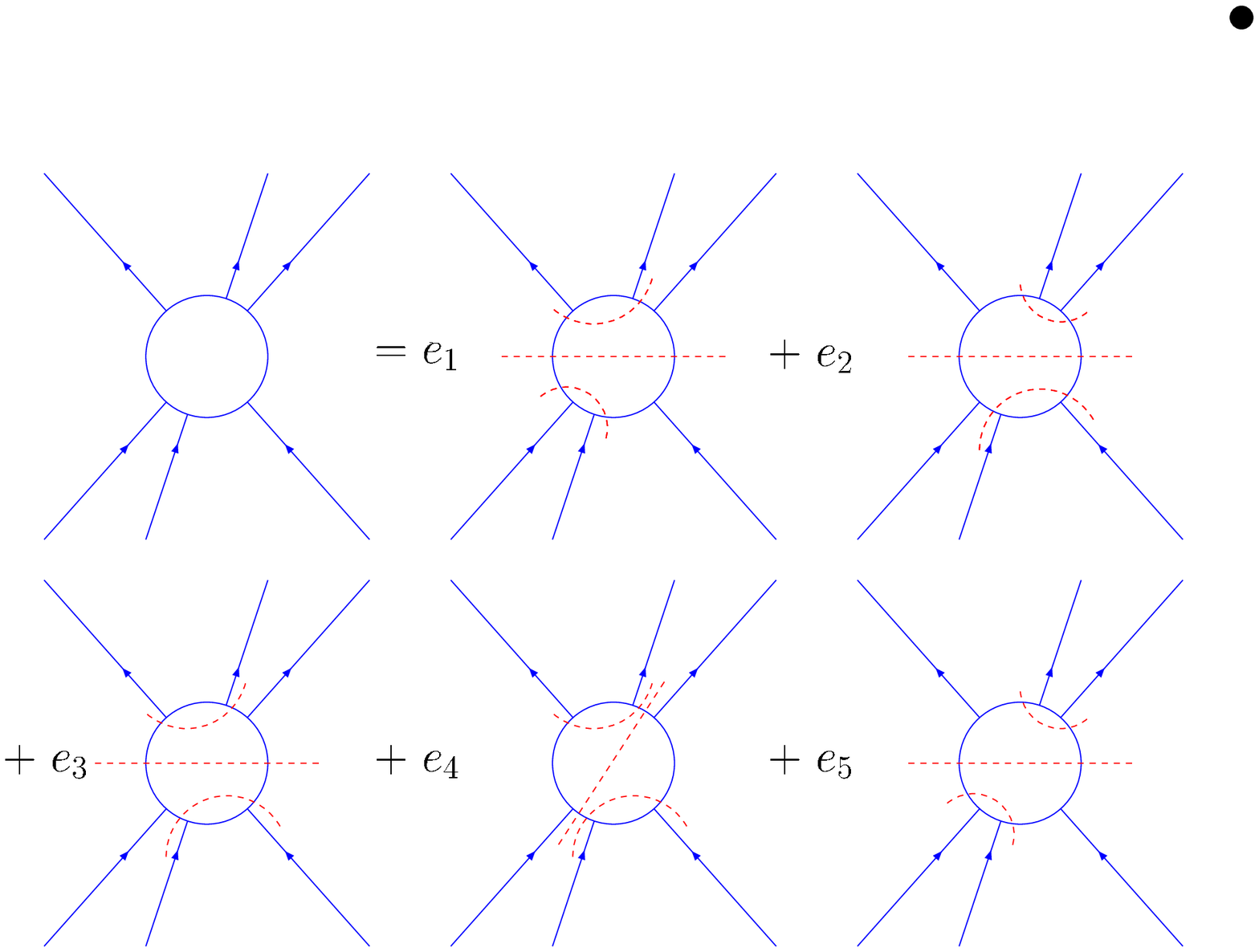,width=11cm,angle=0,bbllx=0,bblly=329,bburx=580,
        bbury=700,clip=}}
\caption{Analytic representation of the amplitude $M_{3\rightarrow 3}$} 
\label{disper33}
\end{figure}

In our previous paper we have studied the high energy Regge behavior of 
scattering amplitudes of $N=4$ supersymmetric Yang Mills theories. In the 
leading logarithmic approximation (LLA) we can make use of the QCD 
calculations since the 
supersymmetric partners of quarks and gluons do not contribute (in this 
limit $t$ channel exchanges with the highest spin dominate).
We now summarize the main results of~\cite{Bartels:2008ce}. For $n$-point 
amplitudes with $n>4$, the high energy scattering amplitudes 
can be written as sums of separate pieces 
(named `analytic representation' or `dispersion representation').
This decomposition reflects the analytic structure required by the Steinmann 
relations~\cite{Steinmann}. The different terms appearing in this 
representation 
can be computed from single energy discontinuities (`imaginary parts') 
or multiple energy discontinuities.
To be definite, we consider the $2 \to 4$ and the $3 \to 3$ scattering 
amplitudes, illustrated in Figs.~\ref{24kinematics} and~\ref{33kinematics}. 
For the $2 \to 4$ scattering         
we are interested in the kinematic limit (double Regge limit) 
\beq
s \gg s_1,s_2,s_3 \gg t_1,t_2,t_3,
\eeq
whereas the $3 \to 3$ scattering process will be studied in the 
limit
\beq
s \gg s_{13},s_{02} \gg s_1,s_3,t'_2=(p_A-p_{A'} -k_2)^2 \gg t_1,t_2,t_3.
\eeq
The analytic decompositions are illustrated in Figs.~\ref{disper24} 
and~\ref{disper33}.

\begin{figure}
\centerline{\epsfig{file=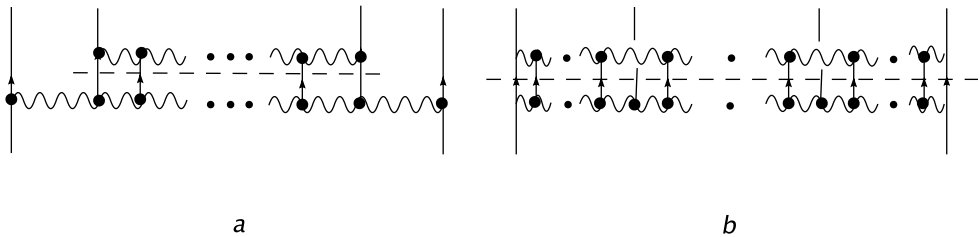,width=13cm,
angle=0,bbllx=94,bblly=498,bburx=524,bbury=606,clip=}}
\caption{(a) the $s_2$ discontinuity for the $2\to4$ amplitude; 
(b) the $s$ discontinuity for the $2\to4$ amplitude} 
\label{24discontinuities}
\end{figure}

In the physical region where all energies are positive, there are, both 
for the $2\to4$ and for the $3\to3$ case, substantial cancellations 
of the Regge cut contributions
between these five terms: in the sum their imaginary parts cancel, and 
the amplitude takes the well-known factorized Regge form. 
In other physical regions, however, where some energies are positive 
and others are negative, the cancellations are less complete, 
and pieces become visible which 
do not show up in the region of only positive energies.
For the $2\to4$ case, the physical region of interest is
\beq
\label{2to4mixed}
s>0,\,\,s_2>0,\hspace{1cm} s_1<0,\,\,s_3<0,\,\,s_{012}<0\,\,,s_{123}<0.
\eeq           
Here non-vanishing discontinuities are only in $s$ and $s_2$, and both 
of them contain a new term which violates the simple factorizing form.
It contains a Regge-cut structure which is described in terms of the color 
octet BFKL equation. We illustrate these discontinuities in 
Fig.~\ref{24discontinuities}.

For the $3 \to 3$ scattering amplitude the corresponding region is 
\beq
\label{3to3mixed}
s>0,\,\,t'_2>0,\hspace{1cm} s_1<0,\,\,s_3<0,\,\,s_{13}<0\,\,,s_{02}<0.
\eeq    
The non-vanishing discontinuities belong to $s$ and $t'_2>0$, and 
they, again, contain the Regge cut pieces. 

In~\cite{Bartels:2008ce} we have compared these results with the 
expression given by Bern et al. 
Whereas for the $2 \to 2$ and $2 \to 3$ amplitudes the QCD results are 
in full agreement with the BDS formula, the $2 \to 4$ and $3 \to 3$ BDS 
amplitudes are correct only in the one loop approximation.
For two or more loops, the Regge cut piece cannot be reproduced by the 
BDS expression.
As explained above, this implies that the BDS formula (in LLA) still gives 
the correct result in the physical region where all energies are positive, but 
it fails (beyond one loop) in the regions (\ref{2to4mixed}) and 
(\ref{3to3mixed}).

\begin{figure}
\centerline{\epsfig{file=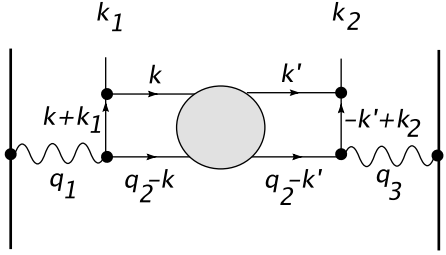,width=6cm,
angle=0,bbllx=183,bblly=526,bburx=380,bbury=636,clip=}}
\caption{The $s_2$ discontinuity for the $2\to4$ amplitude. The big 
blob denotes the BFKL gluon Green's function in colour octet state including 
the reggeization of the lines with momentum $k,q_2-k,k'$ and $q_2-k'$. The 
dots indicate the production vertices in eqs.~(\ref{prodvertex}) 
and~(\ref{phi2}). This is equivalent to Fig.~\ref{24discontinuities}~(a)  
indicating the momentum notation. } 
\label{24discontinuityA}
\end{figure}

In the following we shall investigate these Regge cut pieces in more detail.
Rather than returning to the five terms illustrated in Figs.~3 and 4, 
we directly present an explicit Feynman diagram calculation of the 
single energy discontinuities in $s_2$ and $s$ for the $2 \to 4$
amplitude and in $t_2'$ and $s$ for the $3 \to 3$ amplitude. 
Let us begin with the $s_2$ discontinuity illustrated in 
Fig.~\ref{24discontinuityA}.            
Here the blob in the center denotes the BFKL Green's function in the 
color octet channel which sums the $s$-channel emissions in the center of 
Fig.~\ref{24discontinuities}a, and on both sides we have to convolute this 
Green's function with the `impact factors' $\Phi_1$ and $\Phi_2$. Introducing 
complex momenta 
\beq
k=k_x+ik_y,\,\, k^*=k_x-ik_y
\eeq
and making use of the expression for the vertex describing the production 
of a gluon with definite helicity (cf. eq.(6) of~\cite{Bartels:2008ce}):
\beq
\label{effvertex}
C_{\mu}(\bq_2,\bq_1) e_{\mu}^*(\bk_1) = \sqrt{2} \frac{q_2^*q_1}{k_1^*}
\eeq     
we obtain for the production vertex to the left of the Green's function 
\beq
\label{prodvertex}
\sqrt{2} \frac{q_1(q_2-k)^*}{(k+k_1)^*}. 
\eeq
Here we have used that, in Fig.~\ref{24discontinuityA}, the gluon with 
momentum $\bk+ \bk_1$ 
is on shell (we consider the discontinuity in $s_2$), and at the upper vertex 
where the gluon with momentum $\bk$ is attached  the outgoing gluon helicity 
is conserved. Since the scattering amplitude $T_{2\to n}$ 
for the case of the maximal helicity violation (MHV) can be written 
as~\cite{Bern:2005iz}
 
\beq
\label{factor}
T_{2\to n} = T_{2\to n}^{Born} \cdot M_{2\to n},  
\eeq
we will, throughout our paper, consider the factor 
$M_{2\to n}$ only. We, therefore, separate the production 
vertex of the Born approximation and rewrite (\ref{prodvertex}) as: 
\beq
\sqrt{2} \frac{q_1 q_2^*}{k_1^*}\,
\Phi_1(\bk,\bq_2,\bq_1)
\eeq
with the impact factor:
\beqn 
\label{phi1}
\Phi_1(\bk,\bq_2,\bq_1)&=& \frac{k_1^* (q_2-k)^*}{q_2^*(k+k_1)^*}\nonumber \\ 
&=& 1 - \frac{k^*q_1^*}{q_2^*(k+k_1)^*}.
\eeqn
In the following we shall work with this impact factor.
 
\begin{figure}
\centerline{\epsfig{file=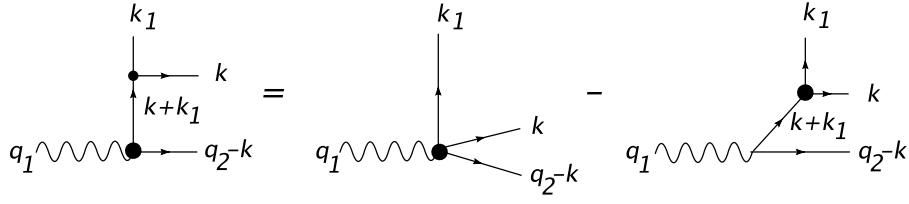,width=12cm,
angle=0,bbllx=88,bblly=539,bburx=496,bbury=632,clip=}}
\caption{Decomposition of the production vertex in eq.(\ref{prodvertex}). The 
dots denote the effective production vertex in eq.(\ref{effvertex}).}
\label{Vertexdecomposition}
\end{figure}

In order to make contact with~\cite{Bartels:2008ce} we should note that, 
with the result in the second line of eq.~(\ref{phi1}), the production vertex 
in (\ref{prodvertex}) can be written as a sum of two terms of the form
\beq
\sqrt{2} \frac{q_1(q_2-k)^*}{(k+k_1)^*}=\sqrt{2} \frac{q_1q_2^*}{k_1^*} 
- \frac{q_1^2}{(k_1+k)^2} \sqrt{2} \frac{(k_1+k)k^*}{k_1^*}.
\label{decomp}
\eeq 
We illustrate this structure in Fig.~\ref{Vertexdecomposition}. The first 
term is `local', {\it i.e.} it has no further dependence on the 
internal momenta, whereas the second one is `nonlocal'.

Similarly, on the right side of the Green's function in 
Fig.~\ref{24discontinuityA} we have 
\beq 
\sqrt{2}\frac{(q_2-k') q_3^*}{k_2-k'}= \sqrt{2} \frac{q_2q_3^*}{k_2}\, 
\Phi_2(\bk',\bq_2,\bq_3)
\eeq
with the impact factor
\beqn
\label{phi2}
\Phi_2(\bk',\bq_2,\bq_3)&=&\frac{k_2(k'-q_2)}{q_2(k'-k_2)}\nonumber\\
& = & 
1-\frac{k'q_3}{(k'-k_2)q_2}
\eeqn
and
\beq
\sqrt{2}\frac{(q_2-k') q_3^*}{k_2-k'} = 
\sqrt{2} \frac{q_2q_3^*}{k_2} + \frac{q_3^2 k'}{(k_2-k')^2} \sqrt{2} 
\frac{(k_2-k')^* }{k_2}.
\eeq 

The discontinuity in $s_2$, to an arbitrary loop accuracy, of the amplitude 
$M_{2\rightarrow 4}$ in LLA then has the form: 
\beq
\label{s2disc}
\frac{1}{\pi}\,\Im_{s_2} \,M_{2\rightarrow 4}=\int _{\sigma -i\infty}^{\sigma +i\infty}
\frac{d\omega}{2\pi i} \left(\frac{s_2}{\mu ^2}\right)^\omega \,f_2(\omega )\,,
\eeq
where the $t_2$-channel partial wave is
\beq
\label{partialwave}
f_2(\omega )=\hat{\alpha}_\epsilon \,
\bq_2^2\int 
d^{2-2\epsilon}k\,d^{2-2\epsilon}k'\,
\Phi _1 (\bk,\bq_2,\bq_1)\,
G^{(8_A)}_{\omega}(\bk,\bk',\bq_2)\,\Phi_2 (\bk',\bq_2,\bq_3)
\eeq
and 
\beq
\hat{\alpha}_\epsilon =\frac{\alpha _s N_c\mu ^{2\epsilon}}
{(2\pi)^{2-2\epsilon}}\,\,\,\,,a= \frac{\alpha_s N_c}{2\pi}\left( 4\pi 
e^{-\gamma} \right)^{\epsilon}.
\eeq
The overall factor $\bq_2^2$  in front of the integral in 
(\ref{partialwave}) takes into account that the
Born approximation of the amplitude contains the pole $1/|q_2|^2$ but in our 
calculations we are interested in the scattering amplitude $M_{2\rightarrow 4}$
with the Born factor being removed (cf.~\ref{factor}).
The Green's function $G^{(8_A)}_{\omega}(\bk,\bk',\bq_2)$ satisfies the
BFKL equation for the color octet channel (putting $\epsilon=0$):
\beq
\label{BFKLoctet}
\omega G_{\omega}^{(8_A)}(\bk,\bk',\bq_2) = \frac{
\delta^{(2)}(\bk-\bk')}{\bk^2 (\bk-\bq_2)^2} 
+ \frac{1}{\bk^2 (\bk-\bq_2)^2} \left(K^{(8_A)} 
\otimes G_{\omega}^{(8_A)} \right)
(\bk,\bk',\bq_2),
\eeq
where $K^{(8_A)}$ 
denotes the BFKL kernel in the color octet channel, containing 
both real emission and the gluon trajectory, and the convolution symbol 
stands for $\otimes = \int \frac{d^2 k}{(2\pi)^3}$. Using complex momenta 
the kernel can be written in the form:
\beq
K^{(8_A)}(k,k';q_2) = \delta^{(2)}(k-k') \left( \omega(-|k|^2) + 
\omega(-|q_2-k|^2) 
\right) +  \frac{a}{2} \frac{k^* (q_2-k) k'(q_2-k')^* + c.c.}{|k-k'|^2}\,,
\eeq
where the gluon trajectory is 
\beq
\omega(-k^2) = a \left( \frac{1}{\epsilon} - \ln \frac{k^2}{\mu^2} \right) .
\eeq
In contrast to the color 
singlet BFKL kernel, the color octet kernel is not infrared finite and needs 
to be dimensionally regularized. It is convenient to separate the 
singular pieces by writing the octet kernel as 
\beqn
K^{(8_A)}(\bk,\bk';\bq_2) &=& \delta^{(2)}(\bk-\bk')
\left[\omega(-\bq^2_2) +   
\frac{1}{2} \left( \omega(-\bk^2) + \omega(-(\bq_2-\bk)^2) 
- 2 \omega(-\bq^2_2) \right) \right] \nonumber\\
&&+ \frac{1}{2} K^{(1)}(\bk,\bk';\bq_2)\nonumber\\
&=& \delta^{(2)}(\bk-\bk') \left[\omega(-\bq^2_2) -   
\frac{a}{2} \ln \frac{\bk^2 (\bq_2-\bk)^2}{\bq^2_2 \bq^2_2} \right]+
\frac{1}{2} K^{(1)}(\bk,\bk';\bq_2).
\eeqn
In this expression, $K^{(1)}(\bk,\bk';\bq_2)$  denotes the color 
singlet BFKL kernel, and infrared singularities are contained in the 
trajectory function $\omega(-\bq^2_2)$.

Inserting this form of the octet kernel into (\ref{s2disc}), the 
discontinuity takes the form:
\beq
\label{s2discreduced}
\frac{1}{\pi}\Im _{s_2} M_{2\rightarrow 4}=s_2^{\omega (t_2)}\,
\int _{\sigma -i\infty}^{\sigma +i\infty}
\frac{d\omega}{2\pi i} 
\left(\frac{s_2}{\mu ^2}\right)^\omega \,
\widetilde{f}_2(\omega )
\eeq 
where the reduced partial wave $\widetilde{f}_2(\omega)$ is given by
\beq
\label{f-reduced}
\widetilde{f}_2(\omega )=\hat{\alpha}_\epsilon \,
\bq_2^2\int
d^{2-2\epsilon}k\,d^{2-2\epsilon}k'\,
\Phi _1 (\bk,\bq_2,\bq_1)\,
\widetilde{G}_{\omega}(\bk,\bk',\bq_2)\,
\Phi_2 (\bk',\bq_2,\bq_3)\,.
\eeq
The Green's function $\widetilde{G}_{\omega}(\bk,\bk',\bq_2)$ 
satisfies the BFKL equation (\ref{BFKLoctet}) with the reduced kernel
\beq
\label{reducedkernel}
\tilde{K}(\bk,\bk';\bq_2) = - \delta^{(2)}(\bk-\bk')  
\frac{a}{2} \ln \frac{\bk^2 (\bq_2-\bk)^2}{\bq^2_2 \bq^2_2}+
\frac{1}{2} K^{(1)}(\bk,\bk';\bq_2). 
\eeq
From the explicit form of the function $\widetilde{f}_2(\omega)$ and of the 
impact factors $\Phi_i$ one sees that
there are potential divergences only for $|\bk| \sim |\bk'| \to 0$ (and not 
for $|\bq_2-\bk| \sim |\bq_2-\bk'| \to 0$). 
The one loop contribution to the partial wave, $\tilde{f}_2^{(0)}$, 
takes the form 
\beqn
\label{1loop2to4}
\omega \widetilde{f}_2^{(0)}(\omega )&=&\hat{\alpha}_\epsilon \,
\bq_2^2\int
d^{2-2\epsilon}k\,
\Phi_1 (\bk,\bq_2,\bq_1)\,\frac{1}{\bk^2 (\bq_2-\bk)^2}
\Phi_2 (\bk,\bq_2,\bq_3) \nonumber\\
&=& \frac{a}{2} \left( \ln \frac{\bk_1^2 \bk_2^2}{(\bk_1+\bk_2)^2 \mu^2} 
- \frac{1}{\epsilon}\right).
\eeqn

In our previous paper~\cite{Bartels:2008ce} we isolated the term
which violates the BDS factorization ansatz. This term, named $V_{cut}$, 
is contained in (\ref{partialwave}): in the impact factors $\Phi_1$ and 
$\Phi_2$ one retains only the nonlocal pieces (cf.~(\ref{decomp}) and 
Fig.~\ref{Vertexdecomposition}), and one 
subtracts the Regge pole contribution. The one loop contribution was given 
in eqs.~(94) and (95) of~\cite{Bartels:2008ce}. In the normalization of 
(\ref{partialwave}) it reads:  
\beqn
\label{C-phase}
\hat{\alpha}_\epsilon \,
\bq_2^2\int 
d^{2-2\epsilon}k\,
\frac{k^*q_1^*}{q_2^* (k+k_1)^*}\,\,\,
\frac{1}{\bk^2 (\bq_2 -\bk)^2}\,\,\,
\frac{k q_3}{q_2 (k-k_2)} = 
\frac{a}{2} \left( \ln \frac{\bq_1^2 \bq_3^2}{(\bk_1+\bk_2)^2 \mu^2} 
- \frac{1}{\epsilon}\right)
\eeqn
and it was shown to coincide (apart from an overall factor) 
with the phase factor $C$ in eq.(75) of~\cite{Bartels:2008ce}. 
In this paper we address the discontinuity in $s_2$, for which we do not 
need to split the impact factors into local and nonlocal pieces.
However, for the discontinuity in $s$ we will come back to the result 
(\ref{C-phase}).

In the following sections we will study the reduced partial wave 
(\ref{f-reduced}) and the reduced kernel (\ref{reducedkernel}) 
in some detail. First we will investigate the infrared properties 
and show that the infrared divergence 
is contained only in the one-loop approximation (\ref{1loop2to4}), {\it i.e.} the   
reduced kernel is infrared finite and introduces no further divergences.
This implies that the divergent term $\sim 1/\epsilon$ is not 
renormalized. 
We will then compute explicitly the two loop approximation to the 
reduced partial wave. Finally, we will return to the reduced kernel and 
calculate its eigenfunctions and eigenvalues.

\section{Infrared properties and eigenvalues of the octet kernel}

In this section we concentrate on the infrared properties of the 
octet kernel. By investigating the most singular part of the reduced 
partial wave (\ref{f-reduced}), we find the exact expression for the 
eigenvalues, and we prove that the infrared singularity of the reduced partial 
wave coincides with the $1/\epsilon$ pole of the one loop approximation.   
The exact solution of the color octet BFKL equation will be derived in section 
5, and it allows to find a closed expression for the reduced partial wave. 

The starting point of our further discussion is eq.(\ref{f-reduced}).
During this section we will denote the transverse momenta by $\bp,\bp'$ and 
$p=p_x+ip_y, p^*=p_x-ip_y$.
The Green's function $\widetilde{G}_{\omega}$ satisfies 
the `renormalized' equation
\beq
\label{renormalizedequation}
\omega \widetilde{G}_{\omega}(\bp,\bp',\bq)=
\frac{1}{\bp^2(\bq-\bp)^2}\,\delta ^{2}(\bp-\bp')-a\,
\widetilde{H} \,
\widetilde{G}_{\omega}(\bp,\bp',\bq_2)\,,
\eeq
where
\beq
\widetilde{H}=\ln \frac{|p|^2|q-p|^2}{|q|^2}
+\frac{1/p}{q^*-p^*}\frac{\ln |\rho |^2}{2} \,p(q-p)^*+
\frac{1/p^*}{q-p}\frac{\ln |\rho |^2}{2}\, p^*(q-p)+2\gamma \,.
\eeq

As we stated before, the most interesting region is the infrared 
divergent region $|p| \sim |p'|\ll |q|$. In this asymmetric kinematic 
it is possible to find eigenvalues and eigenfunctions of the reduced 
kernel. First, the expression 
(\ref{f-reduced}) for $\tilde{f}_2$ is simplified:
\beq
\label{freducedinfra}
\widetilde{f}_2(\omega )=\hat{\alpha}_\epsilon \,
\int
d^{2}p\,d^{2}p'\,
\widetilde{g}_{\omega}(\vec{p},\vec{p}'),
\eeq
and $\widetilde{g}_\omega$ satisfies the equation
\beq
\omega \widetilde{g}_{\omega}(\vec{p},\vec{p}')= 
\frac{1}{|p|^2}\,\delta ^{2}(p-p')-a\, 
\widetilde{H} \,
\widetilde{g}_{\omega}(\vec{p},\vec{p}')\,.
\eeq
Here
\beq
\widetilde{H}=\ln |p|^2
+\frac{1}{p}\,\frac{\ln |\rho |^2}{2} \,p+
\frac{1}{p^*}\,\frac{\ln |\rho |^2}{2} \,p^*+2\gamma 
\eeq
and $\gamma = -\psi(1)$ is the Euler constant.

The Hamiltonian for the octet quantum numbers 
has the property of the holomorphic separability
\beq
\widetilde{H}=\widetilde{h}_8+\widetilde{h}^*_8\,,\,\, 
\widetilde{h}_8=\ln p +\frac{\ln \rho }{2}
+\frac{1}{p}\,\frac{\ln \rho }{2} \,p+\gamma \,. 
\eeq
The holomorphic Hamiltonian $h_8$ is slightly different from the 
corresponding Hamiltonian for the singlet case
\beq
h= \frac{h_P}{2}=\ln p    
+\frac{1}{p}\,(\ln \rho  )\,p+\gamma \,.
\eeq
The difference is  also in the normalization
conditions for the wave functions in these two cases
\beq
||\Psi ||^2_8=\int d^2p\, \Psi ^*|p|^2 \Psi \,,\,\,
||\Psi ||^2_{BFKL}=\int d^2p\, \Psi ^*|p|^4 \Psi \,.
\eeq
The eigenfunctions belonging to the
principal series of the unitary representations of
the M\"{o}bius group in the holomorphic subspace 
have the different form
\beq
\Psi _8^{(m)}=p^{-3/2+m} \,,\,\,
\Psi _{BFKL}^{(m)}=p^{-2+m}\,,\,\,m=\frac{1}{2}+i\nu +\frac{n}{2}
\,.
\eeq
The eigenvalue of the total Hamiltonian for the octet case
is given by
\beq
E^{(m,\widetilde{m})}_8=\epsilon ^{(m)}_8+
\epsilon ^{(\widetilde{m})}_8\,,\,\,\epsilon ^{(m)}_8=
\frac{1}{2}\,\psi \left(\frac{3}{2}-m\right)+
\frac{1}{2}\,\psi \left(\frac{1}{2}+m\right)-\psi (1)\,.
\eeq
To verify this result we act on the wave function $f(k)$ with 
amputated propagators with the Hamiltonian regularized by a mass parameter
$\mu^2$
\begin{eqnarray}
\widetilde{H} f &=& \ln \frac{|k|^2}{\mu^2}\,
|k|^{2i\nu} \left(\frac{k}{k^*}\right)^{n/2}
-\int \frac{d^2k'}{2\pi |k'|^2}\,
\frac{kk^{\prime *}+k^*k'}{
\left(|k-k'|^2+\mu^2\right)}\,|k'|^{2i\nu} 
\left(\frac{k'}{k^{\prime *}}
\right)^{\frac{n}{2}} \nonumber\\
&=&\ln \frac{|k|^2}{\mu^2}\,
|k|^{2i\nu} \left(\frac{k}{k^*}\right)^{\frac{n}{2}}
-\int _0^1dx\int \frac{d^2k'\,(1-x)^{-i\nu +\frac{n}{2}}\,
(1-i\nu +\frac{n}{2})\left(kk^{\prime *}+
k^*k'\right)k^{\prime n}
}{2\pi \left(
|k'-xk|^2+x(1-x)|k^2|+x\mu^2\right)^{2-i\nu +\frac{n}{2}}} \nonumber\\
&=&\left(\ln \frac{|k|^2}{\mu^2}-\frac{\frac{n}{2}}{\nu ^2+
\frac{n^2}{4}}-\int _0^1
\frac{dx\,(1-x)^{-i\nu +\frac{n}{2}}
x^{i\nu +\frac{n}{2}}}{\left(1-x+
\frac{\mu^2}{|k|^2}\right)^{1-i\nu +\frac{n}{2}}}
\right)|k|^{2i\nu} \left(\frac{k}{k^*}\right)^{n/2}.
\end{eqnarray}
One immediately sees that the result is finite when $\mu^2$ is 
taken to zero.   
    
\begin{figure}[ht]
\centerline{\epsfig{file=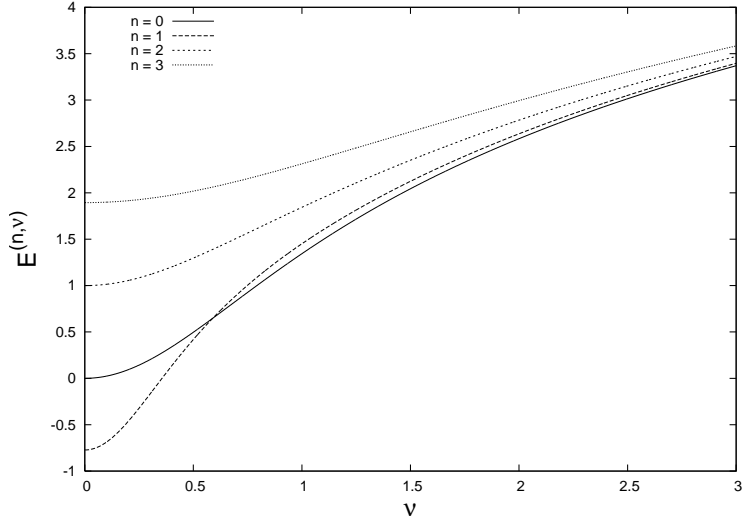,width=7cm,
angle=-90,bbllx=50,bblly=49,bburx=555,
        bbury=770,clip=}}
\caption{Spectrum of the lowest eigenvalues} 
\label{spectrum}
\end{figure}

From this expression we can obtain the eigenvalue
\begin{eqnarray}
E_{\nu n} &=& \frac{1}{2}\left(\frac{1}{i\nu -\frac{|n|}{2}}-
\frac{1}{i\nu +\frac{|n|}{2}}\right)+\psi (1+i\nu +|n|/2)+
\psi (1-i\nu +|n|/2)-2\psi (1) \nonumber\\
&=& \Re \,\psi (1+i\nu +n/2)+\Re \,\psi (1+i\nu -n/2)
-2\psi (1)\,.
\end{eqnarray}
In particular, for $n=0,1$, we have
\begin{eqnarray}
E_{\nu 0} &=& 2\Re \,\psi (1+i\nu )-2\psi (1), \nonumber\\
E_{\nu 1} &=& \frac{1}{2}\frac{1}{\nu ^2+\frac{1}{4}}
+2\Re \,\psi (\frac{1}{2}+i\nu )-2\psi (1)\,.
\end{eqnarray}
The corresponding lowest energies are $E^{(0)}=0$
and $E^{(1)}=2-4\ln 2 <0$\,. In Fig.~\ref{spectrum} we show the 
$\nu$-dependent eigenvalues for different values of $n$. Thus, 
the ground state energy corresponds to $|n|=1$, as it was in the case of
the colorless Odderon state~\cite{NewOdd}. In section 5 we will re-derive this
spectrum by solving the eigenvalue problem exactly.   

The solution of the equation for the Green's function 
can be found with the use of the completeness condition for the
eigenfunctions
\beq
\widetilde{g}_{\omega}(\vec{p},\vec{p}')=\frac{1}{|p|^2|p'|^2}
\,\frac{1}{\pi}\sum _{n=-\infty}^{\infty}e^{i\,n\,(\phi -\phi ')}
\int _{-\infty}^{\infty}
\frac{d\nu}{2\pi} \left(\frac{|p|^2}{|p'|^2}\right)^{i\nu}
\frac{1}{\omega -\omega(\nu , n)} \,.
\eeq
Here
\beq
\omega (\nu ,n)= - \frac{g^2\,N_c}{8\pi ^2}\, E_{\nu n} =  
\frac{g^2\,N_c}{8\pi ^2}\,\left(
2\psi (1)-\Re \,\psi \left(1+i\nu +\frac{n}{2}\right)
- \Re \psi \left(1+i\nu -\frac{n}{2}\right)\right).
\eeq
Inserting this expression into (\ref{freducedinfra}) one sees that 
$\tilde{f}_2$, in this approximation, reduces to the $1/\epsilon$-pole of 
the one-loop expression: the reduced color octet BFKL Green's function 
only leads to finite corrections to the one-loop result and introduces 
no further infrared divergences. Therefore, the divergent contribution is
\beq
\frac{1}{\pi}\Im _{s_2}M_{2\rightarrow 4}|_{div}=
-\frac{a}{\epsilon}\,s_2^{\omega (t_2)+\omega _n(0,0)}=
-\frac{a}{\epsilon}\,s_2^{\omega (t_2)}\,,
\eeq  
where $\omega(0,0)$ is the leading eigenvalue of the `reduced'
octet BFKL kernel discussed before (note that, in this 'infrared' approximation, 
the impact factors are equal to 1, and the solution belonging to the 
eigenvalue $\omega(0,1)$ does not contribute). 
Thus, the divergent contribution $\sim 1/\epsilon$ is not renormalized:
its asymptotic behavior corresponds to the usual gluon Regge pole.  
The reason for this is that the collinear and infrared divergences are 
factorized and that the BDS ansatz is valid in the one-loop approximation.

\section{Contributions at two loops in the perturbative expansion}

Let us return to the reduced partial wave $\tilde{f}_2(\omega)$
in eq.(\ref{f-reduced}) and compute the first terms in the perturbative 
expansion. The one loop 
approximation has already been given in (\ref{1loop2to4}).
As mentioned before, it contains the infrared singularity coming 
from the region  $|k| \to 0$. In this section we iterate the integral 
equation for the Green's function (\ref{renormalizedequation}) inside the partial wave 
and, using the calculus of complex momenta, compute the two loop 
expression. Starting, in Fig.~\ref{24discontinuityA}, 
from the impact factor on the right, given in eq.(\ref{phi2}), 
we have the following expression for the first iteration
(using the reduced color octet BFKL Hamiltonian in eq.(\ref{reducedkernel})):
\begin{eqnarray}
\widetilde{H}\Phi _2 &=& -a\, \ln \frac{|k|^2|q_2-k|^2}{|q_2|^2\,\mu^2}\,
\Phi _2 (\bk,\bq_2,\bq_3) \nonumber\\
&+& a\, \int \frac{d^2k'}{2\pi}\,
\frac{kk^{\prime *}(q_2^*-k^*)(q_2-k')+
k^*k'(q_2-k)(q_2^*-k^{\prime *})}{
\left(|k-k'|^2+\mu^2\right)\,|k'|^2\,|q_2-k'|^2}\,\Phi _2 (\bk',\bq_2,\bq_3)\,.
\end{eqnarray}
Here $\mu^2$ plays the r{\^ o}le of an intermediate infrared cut-off which 
will be removed at the end of our calculations.
The result of integration can be written in the form
\beq
\widetilde{H}\Phi _2=\frac{k_2}{q_2}\,\frac{1}{2}\,\chi (\bk)\,,
\eeq
where 
\beq
\chi (\bk)=-a\, \left( \frac{q_2-k}{k_2-k}\,\ln 
\frac{|k|^2|q_2-k|^2|k_2|^2}{|q_2|^4|k-k_2|^2}
+\frac{q_2}{k_2}\,\ln \frac{|q_2|^2}{|k|^2}
+\frac{(q_2-k_2)k}{k_2(k_2-k)}\,\ln \frac{|q_2-k_2|^2}{|k_2-k|^2}\right) \,.
\eeq
Next we perform the integration over $k$ with the impact factor on the right 
hand side in eq.(\ref{phi1}), and we obtain for the two-loop approximation of the imaginary part  
in the $s_2$-channel in (\ref{s2discreduced}):
\beqn
A_{s_2}&=& \int _{\sigma -i\infty}^{\sigma +i\infty}
\frac{d\omega}{2\pi i} 
\left(\frac{s_2}{\mu ^2}\right)^\omega \,
\widetilde{f}_2(\omega )\nonumber\\
&=&
-\frac{\pi}{2}\bq_2^2\,\, \frac{k_2}{q_2}\,\,a\,\,\ln s_2 \, \int \frac{d^2k}{2\pi \,|k|^2|q_2-k|^2}\,
\Phi _1(\bk,\bq_2,\bq_1)\,\chi (\bk)\,.  
\eeqn
With the use of complex number algebra this two-loop expression can be reduced 
to the form
\beq
A_{s_2}=-\frac{\pi}{2} a^2 \,\ln s_2 \,\int \frac{d^2k}{2\pi}\,\rho (\bk)\,,
\eeq
where
\begin{eqnarray}
\rho (\bk) &=& 
\left(\frac{1}{|k|^2}-\frac{1}{k(k^*+k_1^*)}\right)
\ln \frac{|q_2-k|^2|k_2|^2}{|q_2|^2|k-k_2|^2} \nonumber\\
&+&\left(\frac{1}{k-q_2}\frac{1}{k^*}-\frac{1}{k-q_2}\frac{1}{k^*+k_1^*}\right)
\ln \frac{|q_2-k_2|^2|k|^2}{|q_2|^2|k-k_2|^2} \nonumber\\
&+&\left(\frac{1}{k-k_2}\frac{1}{k^*+k_1^*}-\frac{1}{k-k_2}\frac{1}{k^*}\right)
\ln \frac{|q_2-k|^2|k|^2|q_2-k_2|^2|k_2|^2}{|q_2|^4|k-k_2|^4}\,.
\end{eqnarray}

One can easily verify that the ultraviolet divergences cancel.
Also, in agreement with the previous sections, the divergence at $k=0$
is absent. The above integrals over $k$ can be expressed (with the shift
$k\rightarrow k+c$) in terms of the following expression:
\beq
f(\ba, \bb)\equiv \int \frac{d^2k}{\pi (k-a)(k^*-b^*)}
\ln |k|^2\,. 
\eeq 
To regularize the ultraviolet divergence we introduce the cut-off
\beq
|k|^2<\Lambda ^2\,,
\eeq 
which at the end cancels in the expression for $A_{s_2}$.
One can then write $f$ in the form
\beq
f(\ba, \bb)=\frac{\ln^2 \Lambda ^2}{2}+
f_{r} (\ba, \bb)
\eeq
and use further the regularized value 
$f_{r}$ because
$\ln ^2\Lambda$ is canceled in the final result. 
To calculate this function we take derivatives in the complex coordinates 
$a^*$ and $b$
\beq
\frac{\partial }{\partial a^*}f=-
\frac{1}{a^*-b^*}\ln |a|^2\,,\,\,\frac{\partial }{\partial b}f=-
\frac{1}{b-a}\ln |b|^2\,.
\eeq
After integrating these expressions we obtain
\begin{eqnarray}
f_{r}(\ba,\bb) &=& 
-\int _0^1\frac{dx}{x-\frac{b^*}{a^*}}
\ln x
-\int _0^1\frac{dy}{y-\frac{a}{b}}
\ln y \nonumber\\
&+&\ln |a|^2\,\ln \frac{b^*}{a^*-
b^*}+\ln |b|^2\,\ln \frac{a}{b- 
a}-\frac{1}{2}\ln ^2(a\,b^*)\,.
\end{eqnarray}
The last term was obtained as an integration constant: it 
can be determined from the conditions that it must depend upon  
$a$ and $b*$, and the full function $f$ should depend on the invariants
\beq
|a|^2=a\,a^*\,,\,\,|b|^2=b\,b^*\,,\,\,
a\,b^*= \ba \bb-i[\ba,\bb]_3\,,\,\,
a^*\,b=\frac{|a|^2|b|^2}{a\,b^*}\,.
\eeq
Moreover, from dimensional considerations it follows that it contains
the term $\frac{1}{2}\ln ^2s$, where the invariant $s$ has the 
dimension of $\Lambda ^2$.

The function $f_r$ can be expressed in terms of the Spence's function 
(dilogarithm)
\begin{eqnarray}  
f_{r}(\ba,\bb) &=& 
-Li_2\left(\frac{a^*}{b^*}\right)-Li_2\left(\frac{b}{a}\right)-
\frac{1}{2}\ln ^2(a\,b^*) \nonumber\\
&+& \ln |a|^2\,\ln \frac{b^*}{a^*-b^*}+\ln |b|^2\,\ln \frac{a}{b-a}\,,
\end{eqnarray}
where
\beq
Li_2(z)=-\int_0^z \frac{\ln (1-t)}{t}\,dt\,.
\eeq

Note that the above expression for $f$ has the following  properties
\beq
f(-\ba,-\bb)=f(\ba,\bb)\,,\,\,
f^*(\ba,\bb)=f(\bb,\ba)\,.
\eeq
In some particular cases it can be simplified. For example,
\beq
f_{r}(0, \bb)=\int \frac{d^2k}{\pi k\,(k^*-b^*) } 
\,\ln |k|^2 =- \frac{1}{2}\,\ln ^2(|b|^2)\,.
\eeq

We shall use also the values of the integrals
\beq
\int \frac{d^2k}{\pi |k|^2}\,\ln \frac{|k-c|^2}{|c|^2}=
\frac{1}{2}\,\ln ^2\frac{\Lambda ^2}{|c|^2}\,,\,\,
\int \frac{d^2k}{\pi (k-a)(k-b^*)}=\ln \frac{\Lambda ^2}{|a-b|^2}\,. 
\eeq

With these results we can calculate the two-loop contribution to the 
imaginary part of the amplitude in the $s_2$-channel:
\begin{eqnarray}
-\frac{4}{a^2\ln s_2}\frac{A_{s_2}}{\pi}&=&
\ln \frac{|k_1|^2}{|q_2|^2} \,\ln \frac{|k_2|^2}{|q_2|^2}-
f_{r}(-\bq_2,-\bq_2-\bk_1)+
f_{r}(-\bk_2,-\bk_1-\bk_2) \nonumber\\
&&\hspace{-1cm}
+\ln \frac{|q_2+k_1|^2}{|q_2|^2}\,\ln \frac{|q_2-k_2|^2}{|q_2|^2}
+f_{r}(\bq_2,0)-f_{r}(\bq_2,-\bk_1 )+
f_{r}(\bq_2-\bk_2,-\bk_1-\bk_2) \nonumber\\
&&\hspace{-1cm}
-f_{r}(\bq_2-\bk_2,-\bk_2)
+\ln \frac{|k_2|^2}{|k_1+k_2|^2}\,\ln \frac{|q_2-k_2|^2|k_2|^2}{|q_2|^4}
+f_{r}(\bk_2,-\bk_1)-f_{r}(\bk_2,0 ) \nonumber\\
&&\hspace{-1cm}
+f_{r}(\bk_2-\bq_2,-\bk_1-\bq_2)-f_{r}(\bk_2-\bq_2,-\bq_2)-
2f_{r}(0,-\bk_1-\bk_2)+2f_{r}(0,-\bk_2)\,.
\end{eqnarray}
Using the following properties of dilogarithms
\begin{eqnarray}
Li_2\left(\frac{1}{z}\right) &=& -Li_2(z)-\frac{1}{2}\,\ln ^2(-z)-\zeta _2\,,
\nonumber\\
Li_2\left(1-z\right) &=& -Li_2(z)-\ln (1-z)\,\ln z+\zeta _2\,,
\nonumber\\  
Li_2\left(\frac{z}{1-z}\right) &=& -Li_2(z)-\frac{1}{2}\,\ln ^2(1-z)\,,
\nonumber\\
Li_2\left(\frac{1}{1-z}\right) &=& Li_2(z)+\ln (1-z)\,\ln (-z)-\frac{1}{2}\,
\ln ^2(1-z)+\zeta _2\,,
\nonumber\\  
Li_2\left(\frac{z-1}{z}\right) &=& Li_2(z)+\ln (1-z)\,\ln z-\frac{1}{2}\,
\ln ^2z-\zeta _2\,,
\end{eqnarray}
we can simplify the following sums entering in $2A_{s_2}$
\begin{eqnarray}
-f_r(\bq_2,\bq_2+\bk_1)-f_r(\bq_2,-\bk_1) &=&
\ln |k_1|^2\,\ln |q_1|^2-2 \zeta _2\,,
\nonumber\\
f_r(\bk_2,\bk_1+\bk_2)+f_r(\bk_2,-\bk_1)
&=& -\ln |k_1|^2\,\ln |q_1-q_3|^2+ 2 \zeta _2\,,
\nonumber\\
f_r(\bq_3,-\bk_1-\bk_2)+f_r(\bq_3,\bq_1)
&=&-\ln |q_1|^2\,\ln |q_1-q_3|^2+ 2 \zeta _2\,,
\nonumber\\
-f_r(\bq_3,-\bk_2)-f_r(\bq_3,\bq_2)
&=&\ln |k_2|^2\,\ln |q_2|^2- 2 \zeta _2\,.
\label{sums}
\end{eqnarray}

The final result for $A_{s_2}$ can be written in the very simple form
\beq
\label{2loop2t04}
A_{s_2}=- \pi \frac{a^2}{4}\,\ln s_2 \,\ln \frac{|q_1-q_3|^2|q_2|^2}{|q_1|^2|k_2|^2}\,
\ln \frac{|q_1-q_3|^2|q_2|^2}{|q_3|^2|k_1|^2}\,.
\eeq
It is symmetric with respect to the simultaneous substitutions
\beq
\bk_1 \leftrightarrow \bk_2\,,\,\,
\bq_1 \leftrightarrow -\bq_3\,.
\eeq

In a similar way we can calculate the discontinuity in the $s$-channel. 
Starting, in Fig.~\ref{24discontinuities}b, from the gluon ladders 
in the $t_1$ and the $t_3$ channels, we invoke the bootstrap equation. This equation 
allows us to write, instead of the gluon ladders, simple Regge pole exchanges. 
The resulting 
$s$-discontinuity has the same form as the $s_2$ discontinuity with 
the impact factors $\Phi_1$ and $\Phi_2$ being replaced by    
\beq
\widetilde{\Phi}_1 =
\frac{k^*}{k^*+k_1^*}\,\frac{q_1^*}{q_2^*}\,,\,\,
\widetilde{\Phi}_2 =
\frac{q_3}{q_2} \frac{k'}{k'-k_2}\,.
\eeq
One easily verifies that these modified impact factors coincide 
with the nonlocal pieces of $\Phi_1$ and $\Phi_2$ in eqs.(\ref{phi1}) 
and (\ref{phi2}). We also note that $A_s$ can be obtained from $A_{s_2}$ by substituting
\beq
k_1\leftrightarrow -q_1\,,\,\,k_2 \leftrightarrow q_3,
\eeq
and by changing, inside Fig.6, the integration variables $k \to q_2 -k$,
$k' \to q_2 - k'$.  In fact, in the two loop approximation, 
$A_s$ coincides with $A_{s_2}$: 
\beq
A_s=- \pi \frac{a^2}{4}\,\ln s_2 \,\ln \frac{|q_1-q_3|^2|q_2|^2}{|q_1|^2|k_2|^2}\,
\ln \frac{|q_1-q_3|^2|q_2|^2}{|q_3|^2|k_1|^2}
\eeq 
due to the energy-momentum conservation
\beq
\bk_1+\bk_2=\bq_1-\bq_3\,.
\eeq
\begin{figure}
\centerline{\epsfig{file=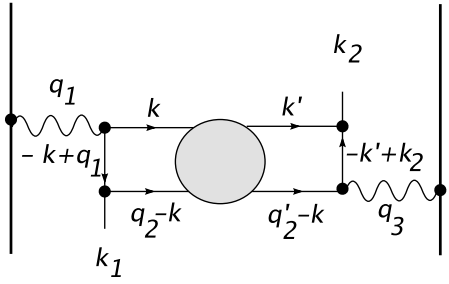,width=6cm,
angle=0,bbllx=186,bblly=526,bburx=380,bbury=649,clip=}}
\caption{The $t'_2$ discontinuity for the $3\to3$ amplitude.} 
\label{33discontinuityB}
\end{figure}

Let us now consider the non-factorisable contribution for the
scattering amplitude $3\rightarrow 3$ (Fig.~\ref{33kinematics}). 
In this case we, again, have the
imaginary parts in $t'_2$ and $s$-channel. For the imaginary part
in the $t'_2$-channel (Fig.~\ref{33discontinuityB}) we have, on the left side, a slightly modified 
impact factor, $\widehat{\Phi}_1$: the 
two corresponding impact factors are
\beq
\widehat{\Phi} _1(\bk,\bq_2,\bq_1)=-\frac{k^*}{k^*-q_1^*}\,
\frac{k_1^*}{q_2^*}\,,\,\,
\Phi_2(k)=
\frac{k'-q_2}{k'-k_2}\,\frac{k_2}{q_2}
\eeq
and, therefore, the infrared divergence at $k=0$ is absent. 
For completeness, we first list the one loop results. 
The one loop result for the partial wave (analogous to 
(\ref{1loop2to4})) is: 
\beq 
\hat{\alpha}_{\epsilon} \bq_2^2 \int d^{2 - 2\epsilon}k \,\,\,
\widehat{\Phi}_1(\bk,\bq_2,\bq_1)  
       \frac{1}{k^2 (q_2-k)^2}
      \Phi_2(\bk,\bq_2,\bq_3)
  = \frac{a}{2} \ln \frac{q_2^2 (q_1+q_3 -q_2)^2}{q_1^2 q_3^2}.
\eeq
Note that, in contrast to the $2 \to 4$ amplitude, there is no infrared 
divergence. Next we quote the 
term analogous to (\ref{C-phase}) which is obtained by retaining, 
in the impact factors, only the `nonlocal' pieces:
\beq 
\hat{\alpha}_{\epsilon} \bq_2^2 \int d^{2 - 2\epsilon}k \,\,\,\frac{q_1^* (q_2 -k)^*}{q_2^* (q_1-k)^*} 
    \frac{1}{k^2 (q_2-k)^2} 
    \frac{k q_3}{q_2 (k-q_2)} = \frac{a}{2} \ln \frac{q_2^2 (q_1+q_3 -q_2)^2}{k_1^2 k_2^2}. 
\eeq
It coincides (up to an overall factor) with the exponent of $C'$ 
in~\cite{Bartels:2008ce} (eq.(80)). In contrast to the $2 \to 4$ amplitude,
this one loop result, again, is infrared finite.

For the two loop calculation we proceed in the same way as for the
$2\to4$ case. Using our previous results for the function $\chi(\bk)$   
we obtain
\beq
A_{t'_2}=\bq_2^2\,\,a \,\,\ln t'_2 \,  \int \frac{d^2k}{2\pi \,|k|^2|q_2-k|^2}\,\frac{k_2}{q_2}\,
\widehat{\Phi}_1(\bk,\bq_2,\bq_1)\,\chi (\bk)\,.
\eeq
With the use of complex number algebra it is possible to transform this
expression into the form
\beq 
A_{t'_2}= - \frac{\pi}{2} a^2 \,\, \ln t'_2 \,\int \frac{d^2k}{2\pi}\,\widetilde{\rho} (\bk)\,,
\eeq
where
\begin{eqnarray}
\widetilde{\rho} (\bk) &=&
\left(\frac{1}{k(k^*-q_2^*)}-\frac{1}{k(k^*-q^*_1)}\right)
\ln \frac{|q_2-k|^2|k_2|^2}{|q_2|^2|k-k_2|^2}
\nonumber\\
&+&\left(\frac{1}{|k-q_2|^2}-\frac{1}{k-q_2}\frac{1}{k^*-q^*_1}\right)
\ln \frac{|q_2-k_2|^2|k|^2}{|q_2|^2|k-k_2|^2}
\nonumber\\
&+&\left(\frac{1}{k-k_2}\frac{1}{k^*-q_2^*}-\frac{1}{k-k_2}\frac{1}{k^*-q_1^*}\right)
\ln \frac{|q_2|^4|k-k_2|^4}{|k|^2|k-q_2|^2|k_2|^2|q_2-k_2|^2}\,.
\end{eqnarray}
The integral over $k$ can be expressed in terms of the 
function $f_r(\ba,\bb)$ introduced above:
\begin{eqnarray}
\frac{2}{a^2\,\ln t'_2} A_{t'_2} &=& f_r(-\bk_2,-\bk_2+\bq_1)
-f_r(-\bq_2,-\bq_2+\bq_1)-\ln \frac{|q_2|^2}{|q_1|^2}\,
\ln \frac{|k_2|^2}{|q_2|^2}+f_r(-\bq_2,0)
\nonumber\\
&&\hspace{-2cm}-f_r(-\bk_2,\bq_3)
+f_r(\bq_2-\bk_2,-\bk_2+\bq_1)-
f_r(\bq_2,+\bq_1)+\ln \frac{q_3^2}{q_2^2}\,\ln \frac{q_2^2}{k_1^2} 
\nonumber\\
&&\hspace{-2cm}-\ln \frac{k_2^2 q_3^2}{q_2^4} \ln \frac{(k_2-q_1)^2}{q_3^2}
-2f_r(0,\bq_1-\bk_2)
\nonumber\\
&&\hspace{-2cm}+2f_r(0,\bq_2-\bk_2)+f_r(\bk_2,\bq_1)-f_r(\bk_2,\bq_2)+
f_r(\bk_2-\bq_2,-\bq_2+\bq_1)-f_r(\bk_2-\bq_2,0)\,.
\end{eqnarray}

With the use of the identities for the sums of the functions
$f(\ba, \bb)$ listed in (\ref{sums}), we can significantly simplify $A_{t'_2}$:
\beq
A_{t'_2}=- \frac{\pi}{4} a^2 \,\ln t'_2\,\ln \frac{|q_2-q_1-q_3|^2|q_2|^2}{|k_1|^2|k_2|^2}\,
\ln \frac{|q_2-q_1-q_3|^2|q_2|^2}{|q_3|^2|q_1|^2}\,.
\eeq
Thus, $A_{t'_2}$ is different from $A_{s_2}$ and $A_{s}$ by the
substitution $\bq_1\leftrightarrow -\bk_1$. In fact, one can also
verify that the same result is obtained for the imaginary part in
$s$ for the $3\rightarrow 3$ transitions.
\beq
A_{s}^{3\rightarrow 3}=- \frac{\pi}{4} a^2 \,\ln t'_2\,\ln \frac{|q_2-q_1-q_3|^2|q_2|^2}{|k_1|^2|k_2|^2}\,
\ln \frac{|q_2-q_1-q_3|^2|q_2|^2}{|q_3|^2|q_1|^2}\,.
\eeq 

As indicated before, all the two loop results are infrared finite and, hence, 
do not affect the infrared singularities in the BDS formula. In the next
section we find the explicit solution at all loops.  

\section{Solution of the BFKL equation in the octet channel}

In this section we solve the eigenvalue problem for the reduced color octet
kernel and derive all-order expression for the $2 \to 4$ and $3 \to 3$ 
amplitudes in the leading-log approximation. 
For the eigenvalue problem it is convenient to return to the symmetric 
notations of the momenta $p_1=p,\,p_2=q-p$
and write the homogeneous BFKL equation for the wave function $f$ with the
removed propagators in the octet channel as follows
\beq
E\,f(\vec{p}_1,\vec{p}_2)=\widetilde{H}\,f(\vec{p}_1,\vec{p}_2)\,,
\eeq 
where $\widetilde{H}$ has the holomorphic separability property
\beq
\widetilde{H}=\widetilde{h}+\widetilde{h}^*\,,\,\,\widetilde{h}
=\ln \frac{p_1\,p_2}{q}+
\frac{1}{2}\left(p_1\ln \rho _{12}\,\frac{1}{p_1}+
p_2\ln \rho _{12}\,\frac{1}{p_2}\right)+\gamma \,.
\eeq
With the use of the relations (see \cite{Int1})
\beq
\ln (z^2\partial )=\ln z+\frac{1}{2}\left(\psi (z\partial )+
\psi(-z \partial+1)\right)\,,\,\,
\ln (\partial)=
-\ln z+\frac{1}{2}\left(\psi (z\partial +1)+\psi(-z \partial)\right) 
\eeq
one can transform the holomorphic Hamiltonian to the form
\beq
\widetilde{h}
=-\ln q+
\frac{1}{2}\left(\ln \left(p_1^2\rho _{12}\right)+
\ln \left(p^2_2\rho _{12}\right)\right)+\gamma \,.
\eeq
By introducing the conjugated variables
\beq
y=\frac{p_1}{p_2}\,,\,\,\partial =\frac{\partial}{\partial y}
=-i\,\frac{p_2^2}{q} \,\rho _{12}\,,
\eeq 
$\widetilde{h}$ can be simplified as follows
\beq
\widetilde{h}=\frac{1}{2}\left(\ln (y^2\partial)+\ln \partial 
\right)+\gamma =\frac{1}2 
\left(\psi (y \partial )+\psi (y \partial +1)\right)+\gamma, 
\eeq
where we neglected pure imaginary terms which cancel
in $\widetilde{H}$.

Thus, the solution of the homogeneous BFKL equation in the 
momentum space can be found in the form
\beq
f _{\nu n}(\vec{k}, \vec{q})=
\left(\frac{k}{q-k}\right)^{i\nu+\frac{n}{2}}
\left(\frac{k^*}{q^*-k^*}\right)^{i\nu-\frac{n}{2}}\,.
\eeq
The corresponding energies were calculated above
\beq
E _{\nu n}=\frac{1}{2}\left[\psi \left(i\nu +\frac{n}{2}\right)
+\psi \left(-i\nu -\frac{n}{2}\right)+\psi \left(i\nu -\frac{n}{2}\right)
+\psi \left(-i\nu +\frac{n}{2}\right)\right]- 2\psi(1)\,.
\eeq
The orthogonality condition for the above wave functions is
\beq
\int \frac{d^2k}{\pi |k|^2|q-k|^2}\,f ^* _{\nu 'n'}(\vec{k}, \vec{q})
\,f _{\nu n}(\vec{k}, \vec{q})=2\pi \delta (\nu '-\nu )\,\delta _{n',n}
\,.
\eeq
Their completeness condition can be written as follows
\beq
\sum _{n=-\infty}^\infty \int _{-\infty}^\infty d\nu \, 
f ^*_{\nu 
n}(\vec{k}', \vec{q}')
\,f _{\nu n}(\vec{k}, \vec{q})=
2\pi ^2\delta ^2(k '-k )\,\frac{|k|^2||q-k|^2}{|q|^2}
\,.
\eeq
Therefore the Green's function for the $t$-channel partial waves
is
\beq
G_\omega (\vec{k}, \vec{k}'; \vec{q})=\frac{1}{2\pi ^2}\,
\frac{|q|^2}{|k|^2||q-k|^2}\,
\sum _{n=-\infty}^\infty \int _{-\infty}^\infty d\nu
\,\frac{f ^*_{\nu
n}(\vec{k}', \vec{q}')
\,f _{\nu n}(\vec{k}, \vec{q})}{\omega -\omega (\nu , n)}\,,   
\eeq
where
\beq
\omega (\nu , n)=-\frac{g^2N_c}{8\pi ^2}\,E _{\nu n}\,.
\eeq

With these results we can find explicit expressions for the $s_2$-discontinuity 
of the $2\rightarrow 4$ scattering amplitude and for the $t'_2$-discontinuity 
of the $3\rightarrow 3$ scattering amplitude. Starting from 
eq.(\ref{f-reduced}), we have to convolute the octet channel Green's function 
with the corresponding impact factors. 
Returning to Fig. \ref{24discontinuityA} and to the notation of section 2 
we have to calculate the integral 
\beq
\label{chi2}
\chi _2=\int \frac{d^2k'}{2\pi }\,\frac{|q_2|^2}{|k'|^2|q_2-k'|^2}
\,\left(\frac{q_2-k'}{k'}\right)^{i\nu +\frac{n}{2}}\,
\left(\frac{q_2^*-{k'}^*}{{k'}^*}\right)^{i\nu -\frac{n}{2}}\,
\frac{k_2(k'-q_2)}{(k'-k_2)q_2}.
\eeq
The simplest way to calculate $\chi _2$ is its differentiation
in $k^*_2$ with the subsequent integration, which gives
\beq
\chi _2=-\frac{1}{2}\frac{1}{\left(i\nu -\frac{n}{2}\right)}\,
\left(\frac{q_3^*}{k^*_2}\right)^{i\nu -\frac{n}{2}}\,
\left(\frac{q_3}{k_2}\right)^{i\nu +\frac{n}{2}}.
\eeq
In a similar way the integral over $k$ gives
\beq
\label{chi1}
\chi _1=\frac{1}{2}\frac{1}{\left(i\nu +\frac{n}{2}\right)}\,
\left(-\frac{q_1}{k_1}\right)^{-i\nu -\frac{n}{2}}\,
\left(-\frac{q^*_1}{k^*_1}\right)^{-i\nu +\frac{n}{2}}\,.
\eeq
As a result,  the imaginary part of the production amplitude in $s_2$
for the transition $2\rightarrow 4$ takes the form 
\beq
\label{s2dis2to4allorder}
\frac{1}{\pi} \Im_{s_2}  M_{2\rightarrow 4} = \frac{a}{4 \pi}
 s_{2}^{\omega(t_2)} \sum _{n=-\infty}^\infty (-1)^n Reg _{s_2}
\int _{-\infty}^\infty \frac{d\nu }{\nu ^2+\frac{n^2}{4}}\,
\left(\frac{q_3^*k^*_1}{k^*_2q_1^*}\right)^{i\nu -\frac{n}{2}}\,
\left(\frac{q_3k_1}{k_2q_1}\right)^{i\nu +\frac{n}{2}}\,
s_2^{\omega (\nu , n)}\,.
\eeq
where the regularization refers to the divergence at $\nu =0$, $n=0$. 
which appears only in in the one loop approximation.
In appendix B we compute the one and two loop results (obtained from 
expanding $s_2^{\omega (\nu , n)} = 1 + \ln s_2\omega (\nu , n) )$, and 
verify the agreement with (\ref{1loop2to4}) and (\ref{2loop2t04}):
\beq
\frac{1}{\pi}\Im _{s_2}\,M_{2\rightarrow 4}=\frac{a}{2  }\,
s_2^{\omega (t_2)}\,\left(\ln \frac{|k_1|^2|k_2|^2}{|k_1+k_2|^2\mu ^2}
-\frac{1}{\epsilon}-\frac{a}{2}\ln s_2 \,\ln
\frac{|k_1+k_2|^2|q_2|^2}{|k_2|^2|q_1|^2}\,
\ln \frac{|k_1+k_2|^2|q_2|^2}{|k_1|^2|q_3|^2}\right).
\eeq

In an analogous way we compute the discontinuity in $s$. In (\ref{chi2}) 
we replace the impact factor $\Phi_2$ by $\widetilde{\Phi}_2$ (and similarly for 
$\Phi_1$ in (\ref{chi1})), and proceed in the  
same way as before. The result can be written in the form
\beq
\frac{1}{\pi}\Im _{s}\,M_{2\rightarrow 4}=\frac{a}{4\pi  }\,
s_2^{\omega (t_2)}\sum _{n=-\infty}^\infty (-1)^n Reg _{s}
\int _{-\infty}^\infty \frac{d \nu}{\nu ^2+\frac{n^2}{4}}\,
\left(\frac{q_3^*k_1^*}{k_2^*q_1^*}\right)^{i\nu -\frac{n}{2}}
\left(\frac{q_3k_1}{k_2q_1}\right)^{i\nu +\frac{n}{2}}
s_2^{\omega (\nu ,n)}\,.
\eeq
with the regularization prescription $Reg _{s}$ for the singularity 
at $\nu=0$, $n=0$ which, again, applies to the 
one loop approximation and takes care of the difference between the 
discontinuities in $s_2$ and $s$.

As a result, the production amplitude
$2\rightarrow 4$ in the multi-Regge kinematics with $s,s_2>0$
and $s_1,s_3<0$ in the leading approximation can be written as
follows
\beq
A_{2\rightarrow 4}=A_{2\rightarrow 4}^{BDS}(1+i\Delta _{2\rightarrow 4})\,,
\eeq
where
\beq
\Delta _{2\rightarrow 4}=\frac{a}{2}\sum _{n=-\infty}^\infty (-1)^n
\int _{-\infty}^\infty \frac{d \nu}{\nu ^2+\frac{n^2}{4}}\,
\left(\frac{q_3^*k_1^*}{k_2^*q_1^*}\right)^{i\nu -\frac{n}{2}}
\left(\frac{q_3k_1}{k_2q_1}\right)^{i\nu +\frac{n}{2}}
(s_2^{\omega (\nu ,n)}-1)\,.
\eeq    
has no infrared singularities. We mention that in the 
region  $s,s_2<0$ and $s_1,s_3>0$ the scattering amplitude has the similar 
form
\beq
A_{2\rightarrow 4}=A_{2\rightarrow 4}^{BDS}(1-i\Delta _{2\rightarrow 4})\,.
\eeq
We emphasize that the correction $\Delta_{2 \to 4}$ does not contribute 
outside these physical regions. 

\begin{figure}[ht]
\centerline{\epsfig{file=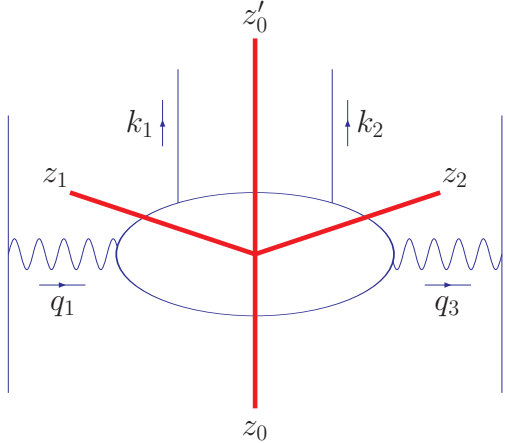,width=8cm,bbllx=10,bblly=412,bburx=400,
        bbury=730,clip=}}
\caption{Dual variables for $A_{2\rightarrow 4}$.} 
\label{dualvariablesfig1}
\end{figure}

It is noteworthy that if we perform the duality transformation shown in 
Fig.~\ref{dualvariablesfig1} (cf. Ref.~\cite{dual})
\beq
q_1\rightarrow z_{01}\,,\,\,k_1\rightarrow z_{0'1}\,,\,\,
q_3\rightarrow z_{02}\,,\,\,k_2\rightarrow z_{20'}\,,
\eeq
and introduce `coordinate' variables $z_i$, 
we see that our results for the imaginary parts depend on the anharmonic ratio
\beq
\label{eq:anhratio}
x=\frac{z_{02}z_{0'1}}{z_{0'2}z_{01}}\,.
\eeq
The reason why the BFKL equation in the octet channel can be solved 
is its invariance under M\"{o}bius transformations in these $z_i$ variables.

It is interesting to note that the correction to the BDS formula in our 
kinematics can be written in terms of four dimensional anharmonic ratios 
~\cite{conformal1,conformal2}. 
In particular, in second order of perturbation theory we can write
\begin{eqnarray}
i \Delta_{2 \to 4}^{(2)} &=& - 2 i \pi \frac{a^2}{4} \ln{s_2} \ln{|k_1+k_2|^2|q_2|^2 \over 
|k_2|^2|q_1|^2} \ln{|k_1+k_2|^2|q_2|^2\over |k_1|^2|q_3|^2} \nonumber \\
&=& \frac{a^2}{4} Li_2(1-\Phi) \ln {(1-\Phi)\over \Phi_2}
\ln{(1-\Phi) \over \Phi_1} + \dots
\end{eqnarray}
where the dots indicate corrections beyond the leading logarithmic accuracy, 
and we have used the notation
\begin{eqnarray}
\Phi = \frac{s s_2}{s_{012} s_{123}},\, 
\Phi_1 = \frac{s_1 t_3}{s_{012} t_2},\,
\Phi_2 = \frac{s_3 t_1}{s_{123} t_2}.  
\end{eqnarray}

An analogous result holds for the $3 \to 3$ amplitudes (for details see 
Appendix B). The discontinuity in $t'_2$ of the scattering amplitude
$3\rightarrow 3$ in the multi-Regge kinematics with $s,t'_2>0$
and $s_1,s_3<0$ in the leading approximation is given by 
\beq
\frac{1}{\pi}\Im _{t'_2}\,M_{3\rightarrow 3}=\frac{a}{4\pi  }\,
t_2^{\prime \,\omega (t_2)}\sum _{n=-\infty}^\infty (-1)^n Reg _{t'_2}
\int _{-\infty}^\infty \frac{d \nu}{\nu ^2+\frac{n^2}{4}}\,
\left(\frac{q_3^*q_1^*}{k_2^*k_1^*}\right)^{i\nu -\frac{n}{2}}
\left(\frac{q_3q_1}{k_2k_1}\right)^{i\nu +\frac{n}{2}}
t_2^{\prime \omega (\nu ,n)}\,,
\eeq
(where, in this case, the regularized integral over $\nu$ for $n=0$ and $a=0$
does not contain any $1/\epsilon$ divergence), 
and the $3 \to 3$ amplitude takes the form 
\beq
A_{3\rightarrow 3}=A_{3\rightarrow 3}^{BDS}(1+i\Delta _{3\rightarrow 3})\,,
\eeq
where
\beq
\Delta _{3\rightarrow 3}=\frac{a}{2}\,
\sum _{n=-\infty}^\infty (-1)^n
\int _{-\infty}^\infty \frac{d \nu}{\nu ^2+\frac{n^2}{4}}\,
\left(\frac{q_3^*q_1^*}{k_2^*k_1^*}\right)^{i\nu -\frac{n}{2}}
\left(\frac{q_3q_1}{k_2k_1}\right)^{i\nu +\frac{n}{2}}
(t_2^{\prime \omega (\nu ,n)}-1)\,.
\eeq
In the region  $s,t'_2<0$ and $s_1,s_3>0$ we can write
\beq
A_{3\rightarrow 3}=A_{3\rightarrow 3}^{BDS}(1-i\Delta _{3\rightarrow 3})\,.
\eeq

\begin{figure}[ht]
\centerline{\epsfig{file=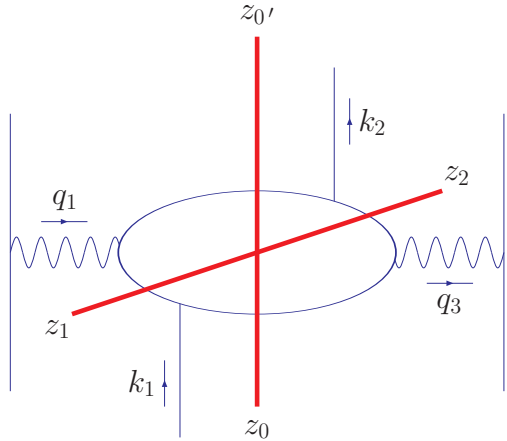,width=8cm,bbllx=10,bblly=412,bburx=400,
        bbury=730,clip=}}
\caption{Dual variables for $A_{3\rightarrow 3}$.} 
\label{dualvariablesfig2}
\end{figure}

Similarly to the $2\to4$ scattering amplitude, if we perform the duality 
transformation (see Fig.~\ref{dualvariablesfig2})
\beq
q_1\rightarrow z_{10'}\,,\,\,k_1\rightarrow z_{01}\,,\,\,
q_3\rightarrow z_{02}\,,\,\,k_2\rightarrow z_{20'}\,,
\eeq
the imaginary parts of the $3 \to 3$ scattering amplitude in the $t'_2$ and 
$s$ channels depend on the same anharmonic ratio (\ref{eq:anhratio}).
Again, the corrections to the BDS formula can be expressed in terms of 
four dimensional anharmonic ratios.

From these results for the $2 \to 4$ and for the $3 \to 3$ amplitudes we conclude that the 
infrared structure of the inelastic amplitudes is given correctly by the 
BDS expression, whereas the finite factors are correct only in the one loop 
approximation.

\section{Conclusions}
In this paper we studied, in the leading logarithmic approximation,
the cut contribution which, in our previous paper, was found to violate 
the simple Regge factorization of the BDS formula. As a main result we have 
verified that the factorization of universal infrared singularities is
 not affected, {\it i.e.}
 the violation of the 
BDS formula is in the finite pieces. We have computed the energy spectrum 
of the color octet BFKL Hamiltonian, and we have concluded that the infrared 
divergent gluon trajectory can be separated from the finite remainder of the 
BFKL Green's function. We have explicitly computed the two loop approximation 
of the Regge cut piece. The integral equation for the wave function 
of two reggeized gluons in
the octet channel is solved explicitly and the intercepts of the Regge
singularities are calculated.\\[1cm]   
{\bf Acknowledgements}: JB is grateful for the hospitality of CERN where 
part of this work has been done, and ASV gratefully acknowledges 
the hospitality of DESY and Hamburg University. LL is supported by the RFBR 
grants 06-02-72041-MNTI-a, 07-02-00902-a and RSGSS-5788.2006.2.

\appendix

\setcounter{equation}{0}

\renewcommand{\theequation}{A.\arabic{equation}}
\section{Forward scattering with color octet exchange}

In this appendix we relate the octet equation in the 
forward case to the BFKL equation for the singlet case. As it will be
shown, its physical solution exists only if the gluon Regge trajectory
is not expanded in a power series on $\epsilon$.

The relevant 
integral equation to be satisfied by the gluon Green's function in the octet 
case reads
\beq
\omega G_{\omega}(\vec{p},\vec{p}',\vec{q})=
\frac{1}{|p|^2|q-p|^2}\,\delta ^{2-2\epsilon }(p-p')-a\,H_\epsilon \,
G_{\omega}(\vec{p},\vec{p}',\vec{q}_2)\,,
\eeq
where
\beq
a=\frac{\alpha _s N_c}{2\pi}\left(4\pi e^{-\gamma}\right)^{\epsilon },
\eeq
and $\gamma$ is the Euler's constant. From now on we shall use the notation 
$\vec{p}_1 \equiv\vec{p}$ and $\vec{p}_2 \equiv \vec{q}-\vec{p}$ for the 
transverse momenta of the two Reggeized gluons.

The BFKL Hamiltonian for the channel with the octet quantum numbers
can be written in operator form as
\beq
H_\epsilon=\ln \frac{|p_1|^2}{\mu ^2}-\frac{1}{\epsilon } +\ln |p_2|^2+
\frac{1}{p_1p^*_2}\frac{\ln |\rho _{12}|^2}{2} p_1p^*_2+
\frac{1}{p^*_1p_2}\frac{\ln |\rho _{12}|^2}{2} p^*_1p_2+2\gamma \,,
\eeq
where we have neglected terms of ${\cal O} (\epsilon)$ and introduced for 
the two Reggeized gluons  the
complex variables $\rho _k,\rho ^*_k$ ($\rho _{12}=\rho _1-\rho _2$) 
and their canonically conjugated momenta $p_k,p^*_k$.
The dependence on the divergence $1/\epsilon$ can be removed by the following 
shift in the parameter $\omega$:
\beq
\omega \rightarrow \omega +\frac{a}{\epsilon}\,,
\eeq
which leads to the appearance of a Sudakov-type infrared 
divergent factor in the amplitude $M_{2\rightarrow 4}$, {\it i.e.} 
\beq
M_{2\rightarrow 4}\rightarrow Z\,M_{2\rightarrow 4}\,,\,\,Z=
\exp \left(\frac{a}{\epsilon}\,\ln \frac{s_2}{\mu ^2}\right)\,.
\label{Z}
\eeq
We can then work with the renormalized Hamiltonian $H$ removing 
the divergent term
\beq
H=H_\epsilon +\frac{1}{\epsilon}\,.
\eeq

It is known that the BFKL equation in the color singlet channel is M\"{o}bius 
invariant in coordinate space. Its solutions are
\beq
E_{m,\widetilde{m}}(\vec{\rho}_{10},\vec{\rho}_{20})=
\left(\frac{\rho _{12}}{\rho _{10}\rho _{20}}\right)^m
\left(\frac{\rho ^*_{12}}{\rho ^*_{10}\rho ^*
_{20}}\right)^{\widetilde{m}}\,,
\label{Greensfunction}
\eeq
where $m,\widetilde{m}$ are conformal weights 
\beq
m=\frac{1}{2}+i\nu +\frac{n}{2}\,,\,\,
\widetilde{m}=\frac{1}{2}+i\nu - \frac{n}{2}
\eeq
and $\vec{\rho} _0$ is the Pomeron coordinate. 
The expression~(\ref{Greensfunction}) corresponds to the three-point Green's 
function with non-amputated legs. The scalar product of two of these 
functions is defined by
\beq
<E_{m,\widetilde{m}}||E_{m',\widetilde{m}'}>=
\int d^2\rho _1\,d^2\rho _2\,
E_{m,\widetilde{m}}(\vec{\rho}_{10},\vec{\rho}_{20})
\Delta _1\,\Delta _2 
E^*_{m',\widetilde{m'}}(\vec{\rho}_{10'},\vec{\rho}_{20'})\,,
\eeq
where $\Delta _k$ are the corresponding Laplace operators.

\subsection{The solution for the octet case at $q_2=0$}

The solution in momentum space for $q_2=0$ in the color singlet case
can be obtained by using the Fourier transform of the singlet 
solution~(\ref{Greensfunction}), {\it i.e.}
\beq
f_{m,\widetilde{m}}(\vec{p} )=\int d^2\rho _{12}\,d^2\rho _0
e^{i\vec{p}\vec{\rho}_{12}}
\,E_{m,\widetilde{m}}(\vec{\rho}_{10},\vec{\rho}_{20}) \sim
p^{m-2}\,(p^*)^{\widetilde{m}-2}\,.
\eeq
It is convenient to introduce the new function $\phi _{\nu ,n}(\vec{p})$ 
as follows
\beq
f_{m,\widetilde{m}}(\vec{p})=|p|^{-3}\phi _{\nu ,n}
(\vec{p})\,,
\eeq  
with normalization
\beq
\phi _{\nu ,n}(\vec{p})=\left(\frac{|p|^2}{\mu ^2}
\right)^{i\nu}e^{i\alpha n}\,,\,\,p=|p|e^{i\alpha}\,.
\eeq
These functions satisfy the following orthonormality
and completeness properties
\begin{eqnarray}
\int \frac{d^2p}{|p|^2}\,\phi _{\nu ,n}(\vec{p})\,
\phi ^*_{\nu ',n'}(\vec{p}) &=& 2\pi ^2
\,\delta (\nu -\nu ')\,\delta _{n,n'}\,,\\
\sum _{n=-\infty}^{\infty}
\int _{-\infty}^{\infty}d\nu \,\phi _{\nu ,n}(\vec{p})
\,\phi ^*_{\nu ,n}(\vec{p}') &=& 4\pi ^2
\,\delta (\ln |p|^2-\ln |p'|^2)\,\delta (\alpha -\alpha ')\,.
\end{eqnarray}
The homogeneous BFKL equation for the octet case can be written in the form 
\beq
\omega \psi =-a H \psi \,,\,\,H=\ln \frac{|p|^2}{\mu ^2}+
\frac{1}{2}H_0\,,
\label{octethamiltonian}
\eeq
where $H_0$ is the Hamiltonian for the singlet case.
Its solution can be constructed in terms of
the linear combination of the functions 
$\phi _{\nu ,n}(\vec{p})$
\beq
\psi _{\omega ,n}(\vec{p})=e^{i\alpha n}\int _{-\infty}^{\infty} d\nu \,
\left(\frac{|p|^2}{\mu ^2}\right)^{i\nu}\,a _{\omega ,n}(\nu )\,.
\label{solt0}
\eeq 
Taking into account~(\ref{octethamiltonian}) the function 
$a _{\omega ,n}(\nu )$ should satisfy the equation
\beq
\omega a _{\omega ,n}(\nu )=-a\left[i\frac{\partial}{\partial \nu}
+\psi \left(\frac{1}{2}+i\nu +\frac{|n|}{2}\right)+
\psi \left(\frac{1}{2}-i\nu +\frac{|n|}{2}\right)
+2\gamma \right] a _{\omega ,n}(\nu ).
\label{eqnu}
\eeq
Its solution is
\beq
a _{\omega ,n}(\nu )=\exp \left[i\nu\,
\left(\frac{\omega}{a}+2\gamma \right)\right]
\frac{\Gamma \left(\frac{1}{2}+i\nu +
\frac{|n|}{2}\right)}{\Gamma \left(\frac{1}{2}-i\nu +\frac{|n|}{2}\right)}\,.
\eeq
For $a _{\omega ,n}(\nu )$ we have the following normalization 
\beq
\int _{-\infty}^{\infty}d\nu \,a _{\omega ,n}(\nu )
\,a^* _{\omega' ,n}(\nu  )=2\pi a\delta (\omega -\omega ')
\eeq
and completeness conditions
\beq                  
\int _{-\infty}^{\infty}d\omega \,a _{\omega ,n}(\nu )
\,a^* _{\omega ,n}(\nu ' )=2\pi a\delta (\nu -\nu ') \,.  
\eeq
Therefore the completeness relation for the eigenfunctions
$\psi _{\omega ,n}(\vec{p})$ has the form
\beq
\sum _{n=-\infty} ^{\infty}  
\int _{-\infty}^{\infty}\frac{d\omega }{2\pi}\,
\psi _{\omega ,n}(\vec{p})
\,\psi ^*_{\omega ,n}(\vec{p'})=a\,(2\pi )^2\,
\delta (\ln |p|^2 -\ln |p'|^2)\,\delta (\alpha -\alpha ')\,.
\eeq
Using these expressions, we can solve the inhomogeneous equation
for the Green's function $g_\omega (\vec{p},\vec{p}',0)$
\beq
\omega g_{\omega}(\vec{p},\vec{p}',0)=
(2\pi )^2\,\frac{a}{2}\,|p|^2\,\delta ^2(p-p')-a\,H \,
g_{\omega}(\vec{p},\vec{p}',0)
\eeq
in terms of a superposition of eigenfunctions of the homogeneous 
equation, {\it i.e.}
\beq
g_{\omega}(\vec{p},\vec{p}',0)=\sum _{n=-\infty} ^{\infty}
P\int _{-\infty}^{\infty}\frac{d\omega '}{2\pi}
\frac{\psi _{\omega' ,n}(\vec{p})
\,\psi ^*_{\omega' ,n}(\vec{p'})}{\omega -\omega '}\,,
\eeq
where $P$ means that the integral over $\omega$ is taken with the
principal value prescription.

In terms of this Green's function the $t_2$-channel partial wave can be 
written as
\beq
f_2(\omega )=\frac{Z}{4\pi} 
\int _0^{\infty}\frac{d|k|^2}{|k|^3}\int _0^{2\pi }  
\frac{d\alpha \,k^*k^*_1}{k^*+k^*_1}
\int _0^{\infty}\frac{d|k'|^2}{|k'|^3}\int _0^{2\pi} 
 \frac{d \alpha' \,k'k_2}{k_2-k'}\,  
g_{\omega}(\vec{k},\vec{k}',0)\,,
\eeq
where $Z$ is the divergent factor discussed in~(\ref{Z}). 
We can now write $f_2(\omega )$ in a different form using the explicit
expression for $g_{\omega}(\vec{k},\vec{k}',0)$:
\beq
f_2(\omega )=\frac{Z}{4\pi}\sum _{n=-\infty} ^{\infty}
P\int _{-\infty}^{\infty}\frac{d\omega '}{2\pi i(\omega -\omega ')}\,
b_{\omega 'n}(\vec{k})
\eeq
where
\beq
b_{\omega n}(\vec{k}_1)=\int _{-\infty}^{\infty} d\nu
\,K_{\nu n}(\vec{k}_1) \,
a_{\omega ,n }(\nu )\,,
\eeq
\beq
\widetilde{b}_{\omega n}(\vec{k}_2)=\int _{-\infty}^{\infty} d\nu 
\,\widetilde{K}_{\nu  n}(\vec{k}_2) \,
a^*_{\omega ,n }(\nu )\,.
\eeq
The functions $K$ and $\widetilde{K}$ read
\beq
K_{\nu n}(\vec{k}_1)=\int _0^{\infty}
\frac{d|k|^2\,k^*_1}{|k|^3}\frac{|k|^{2i\nu}}{\mu ^{2i\nu}}
\int _0^{2\pi} \frac{d\alpha \,k^*e^{i\alpha n}}{k^*+k^*_1}=
\frac{(-1)^{n-1}2\pi \,\Phi _{\nu, |n|}    (\vec{k}_1)
}{\frac{|n|-1}{2}+i\nu }\,, 
\eeq
\beq
\widetilde{K}_{\nu n}(\vec{k}_2)=\int _0^{\infty}
\frac{d|k'|^2 \,k_2}{|k'|^3}\frac{|k'|^{-2i\nu}}{\mu ^{-2i\nu }}
\int _0^{2\pi} \frac{d\alpha' \,k'e^{-i\alpha ' n}}{k_2-k'} 
=\frac{(-2\pi )\,\Phi ^*_{\nu, |n|}(\vec{k}_2)}{\frac{|n|-1}{2}-i\nu } 
\,,
\eeq
where
\beq
\Phi _{\nu, |n|} (\vec{k}_1)=\left(\frac{|k_1|^2}{\mu ^2}
\right) ^{i\nu }\,e^{i\alpha _1(|n|-1)}.
\eeq

Thus, one can obtain the following simple representation for the functions
$b$ and $\widetilde{b}$
\begin{eqnarray}
b_{\omega n}(\vec{k}_1) &=& -2\pi (-1)^n\,e^{i\alpha _1(|n|-1)}
c_{\omega n}(\vec{k}_1)\,,\\
\widetilde{b}_{\omega n}(\vec{k}_2)
&=&-2\pi \,e^{-i\alpha _2(|n|-1)} c^*_{\omega n}(\vec{k}_1)\,,
\end{eqnarray}
where
\beq
c_{\omega n}(\vec{k}_1)=\int _{-\infty}^{\infty} d\nu
\,e^ {i\nu\,
\left(\frac{\omega}{a}+2\gamma +\ln \frac{|k_1|^2}{\mu ^2}\right)}
\,\frac{\Gamma (\frac{|n|-1}{2}+i\nu )}{\Gamma (\frac{|n|+1}{2}-i\nu)}\,.
\eeq

The final result for the imaginary part of the amplitude in the variable
$s_2$ reads
\beq
\frac{\Im _{s_2} M_{2\rightarrow 4}}{\pi}=Z\pi \sum _{n=-\infty} ^{\infty}
(-1)^ne^{i\alpha _{12}(|n|-1)}\,
\int _{-\infty}^{\infty}d\omega \,\left(\frac{s_2}{\mu ^2}\right)^{\omega}\,
c_{\omega n}(\vec{k}_1)\,c^*_{\omega n}(\vec{k}_2)\,,
\eeq
where $\alpha _{12}=\alpha _1-\alpha _2$. There is an ambiguity in the 
integration over $\nu$ at $\nu =0$ for $|n|=1$, but at that point 
$c_{\omega n}(\vec{k}_1)$ does not depend on $\omega$.

\subsection{Spectrum quantization}

It is important to note that in the region $\vec{p}\rightarrow 0$ we can not 
use the simplest form for the Regge trajectory in the Born approximation and 
we should write the exact expression instead: 
\beq
\ln |p|^2 -\frac{1}{\epsilon} \rightarrow E_g (|p|)=
-\frac{1}{\epsilon}\,\left(\frac{|p|^2}{\mu ^2}\right)^{-\epsilon}\,.
\eeq
The reason for this is that the solution $\psi _{\omega,n}(\vec{p})$ has a 
good behavior only for large $|p|$, {\it i.e.}
\beq
\lim _{|p|\rightarrow \infty}\,\psi _{\omega ,n}(\vec{p})
\sim e^{i\alpha 
n}
\,
\left(\frac{|p|^2}{\mu ^2}\right)^{-\frac{1+|n|}{2}}\,.
\eeq
In the region of small $|p|$ its asymptotics is given 
by the saddle point contribution in the integral
over $\nu$ and is not stable for the simplified expression
$E_g (|p|)$ at $\epsilon \rightarrow 0$. The position 
of this saddle point $\nu$ is
defined by the solution of the BFKL equation in the
classical approximation
\beq
\omega =-a\left[E_g (|p|)+\frac{1}{\epsilon}
+\psi \left(\frac{1}{2}+i\nu +\frac{|n|}{2}\right)+
\psi \left(\frac{1}{2}-i\nu +\frac{|n|}{2}\right)+2\gamma \right]\,.
\label{clBFKL}
\eeq
In this expression we have used the exact expression $E_g (|p|)$
for the gluon energy. Let us indicate that, due to the symmetry of
$E_0(\nu )$ under the substitution $\nu \rightarrow -\nu$, there
are two solutions of this equation related by this symmetry and the 
semiclassical expression for $\psi _{\omega ,n}(\vec{p})$ oscillates 
in this region.
  
The intercept $\Delta$ of the corresponding singularity in the
$j-1$-plane of the $t$-channel corresponds to the values 
$\nu =0, n=0$ of the M\"{o}bius parameters 
(for $\epsilon <0$)
\beq
\Delta =-a\min _{|p|}\left(E_g (|p|)
-4\ln 2 \right)=\frac{\omega _{P}}{2}\,,\,\,\omega _{P}=
\frac{g^2}{\pi ^2}\,N_c\,\ln 2\,,
\label{Delt}
\eeq
where $\omega _P$ is the intercept of the BFKL Pomeron.

Let us solve the Schr\"{o}dinger equation for the wave 
function 
with the modified 
expression for the Regge trajectory analytically. For
this purpose we shall use the representation where
the coordinate is
\beq
x=\ln \frac{|p|^2}{\mu ^2}\,.
\eeq
The Schr\"{o}dinger equation has the form
\beq
E_0(\nu ,n)\Psi _{\nu ,n}(x)=\left(-\frac{e^{-\epsilon \,x}}{\epsilon}+
\frac{H_0}{2}\right)\,\Psi _{\nu ,n}(x)\,, 
\eeq
where $\epsilon \rightarrow -0$ and $E_0(\nu, n)$ is the total energy
at $x\rightarrow -\infty$
\beq
E_0(\nu, n)=\psi \left(\frac{1}{2}+i\nu +\frac{|n|}{2}\right)+
\psi \left(\frac{1}{2}-i\nu +\frac{|n|}{2}\right)+2\gamma \,.
\eeq
At $x\rightarrow -\infty$ the potential energy goes to zero and
we can search for two solutions of this equation of the form
\beq
\Psi _{\nu ,n}^{\pm }(x)=e^{\pm i\nu x}\,
\sum _{r=0}^{\infty}C^{\pm}_r(\nu )\,e^{-\epsilon \,r\,x}\,,
\label{expansionpsi}
\eeq
where $C^{\pm}_r{(\nu)}$ satisfies the recurrence relation
\beq
E_0(\nu, n)\,C^{\pm}_r(\nu )=-\frac{1}{\epsilon}\,C^{\pm}_{r-1}(\nu )
+E_0(\nu \pm i\,r, n)\,C^{\pm}_r(\nu ).
\eeq
Therefore one can write the following expression for
the coefficients $C^{\pm}_r(\nu )$
\beq
C^{\pm}_r(\nu )= \left(\frac{-1}{\epsilon}\right)^r\,
\prod _{t=1}^r\frac{1}{E_0(\nu, n)-E_0(\nu \pm i \,t, n)}\,.
\eeq
In principle, the expansion in~(\ref{expansionpsi}) with
these coefficients is convergent for all values of $x$ and therefore we 
could find at least numerically the linear combination of $\Psi ^+$ and 
$\Psi ^-$ for which the wave function decreases at $x \rightarrow \infty$. 

We consider now the case of small $\nu$, where we can use the diffusion 
approximation for $E_0$:  
\beq
E_0(\nu ,n) =2\,\psi \left(\frac{1}{2}+\frac{|n|}{2}\right)+
2\gamma -\psi ''\left(\frac{1}{2}+\frac{|n|}{2}\right)\,\frac{\nu ^2}{2} \,.
\eeq
In particular for $n=0$ we have
\beq
E_0(\nu ,n)=-4\,\ln 2+14\,\zeta (3)\,\nu ^2\,.
\eeq
In this case one obtains
\beq
C^{\pm}_r(\nu )= \left(\frac{-1}{14\,\zeta (3)\, \epsilon}\right)^r\,
\prod _{t=1}^r\frac{1}{t(t\pm 2i\nu )}=
\frac{(-14\,\zeta (3)\,\epsilon )^{-r}\Gamma (1\pm 2i\nu )}{\Gamma (r+1)\,
\Gamma (r+1\pm 2i \nu )}\,.
\eeq
As a result, we can express $\Psi ^{\pm}$ with an appropriate normalization
constant in terms of the Bessel function with imaginary argument (for
$\epsilon <0$)
\beq
\Psi _{\nu ,0}^{\pm }(x)=I_{\pm 2i\nu}
\left(\sqrt{\frac{2\,\exp (-\epsilon x)}{-7\zeta (3)\epsilon}}\right)\,.
\eeq
The solution, which has the good asymptotic behavior at $x \rightarrow \infty$
\beq
\Psi _{\nu ,0}(x) \sim 
e^{-\sqrt{\frac{2\,\exp (-\epsilon x)}{(-7\zeta (3)\epsilon )}}}\,,
\eeq
is 
\beq
\Psi _{\nu ,0}(x)=K_{2i\nu}\left(\sqrt{\frac{2\,
\exp (-\epsilon x)}{-7\zeta (3)\epsilon}}\right)\,,
\eeq
where
\beq
K_{2i\nu}(z)=\frac{\pi}{2}\,
\frac{I_{-2i\nu}(z)-I_{2i\nu}(z)}{\sin (2\pi i\nu)}\,.
\eeq

\renewcommand{\theequation}{B.\arabic{equation}}

\setcounter{equation}{0}

\newpage
\section{Calculation of the one and two loop contributions}
In this appendix we calculate, starting from (\ref{s2dis2to4allorder}),
the one and two loop approximations.
For the one loop approximation we compute the integral: 
\beq
M=\sum _{n=-\infty}^\infty (-1)^n\int _{-\infty}^\infty
\frac{d\nu }{\nu ^2+n^2/4}
\,\beta ^{\,i\nu}
\alpha ^{n/2}\,,
\eeq
where 
\beq
\alpha =\frac{q_3k_1q_1^*k_2^*}{q_3^*k_1^*q_1k_2}
\eeq
and 
\beq
\beta = \frac{|q_3|^2|k_1|^2}{|k_2|^2|q_1|^2}.
\eeq
We begin with the terms $n \neq 0$ and integrate over $\nu$: 
\beqn
M_{n \neq 0}=&=&\theta (\beta -1)\left(\sum  _{n=1}^\infty
\frac{2\pi}{n}(-1)^n\beta ^{-n/2}\left(\alpha ^{n/2}+\alpha ^{-n/2}
\right)\right) \nonumber\\
&&
+\theta (1-\beta )\left(\sum  _{n=1}^\infty
\frac{2\pi}{n}(-1)^n\beta ^{n/2}\left(\alpha ^{n/2}+\alpha ^{-n/2}
\right) \right) \nonumber \\
&=&-\theta (\beta -1)2\pi
\left(\ln \left((1+\sqrt{\alpha}/\sqrt{\beta})
(1+1/\sqrt{\alpha \beta })\right) \right)
\nonumber\\
&&-\theta (1-\beta )2\pi
\left(\ln \left((1+\sqrt{\beta}/\sqrt{\alpha})                      ]
 (1+\sqrt{\alpha \beta })\right) \right)
\eeqn
Using (B.2) and (B.3) we obtain:
\beq
M_{n \neq 0} = -2\pi \left( \ln \frac{|k_1+k_2|^2|q_2|^2}{|k_1k_2q_1q_3|} -\frac{1}{2}|\ln \beta| \right)
\eeq

For the term $n=0$ the divergence at $\nu=0$ needs to be 
regularized. In order to reproduce the one loop result (\ref{1loop2to4}) 
derived in dimensional regularization we need 
\beq
Reg _{s_2}\int _{-\infty}^\infty \frac{d \nu}{\nu ^2}\,
\left|\frac{q_3k_1}{k_2q_1}\right|^{2i\nu}=2\pi \left(-\frac{1}{\epsilon}
+\ln \frac{|q_2|^2}{\mu ^2}- 
\ln \left|\frac{q_3 q_1 }{k_1 k_2}\right|
- \left| \ln \frac{|q_3| |k_1|}{|k_2| |q_1|}\right| \right). 
\eeq
The sum of this contribution for $n=0$ and $M_{n \neq 0}$ 
is (apart from the overall factor) in agreement with the one loop result 
in (\ref{1loop2to4}). 

Next let us consider the two-loop contribution. We need to calculate the 
following integral:
\beq
R=\sum _{n=-\infty}^\infty (-1)^n\int _{-\infty}^\infty
\frac{d\nu \,\,E_{\nu n}}{\nu ^2+n^2/4}
\,\beta ^{\,i\nu} \alpha^{n/2}
\eeq
where according to eq. (38) and (85)
\beq
E_{\nu n}=-\frac{|n|}{\nu ^2+\frac{n^2}{4}}+\sum _{k=0}^\infty
\left(\frac{2}{k+1}-\frac{1}{k+1+i\nu +|n|/2}-
\frac{1}{k+1-i\nu +|n|/2}\right)\,.
\eeq
The integral over $\nu$ can be calculated by residues, and we take into 
account that the contributions from the poles $\nu =\pm i|n|/2$ exist only for 
$n\ne 0$. As for the other poles, they give contributions also at $n=0$. 
Thus, we obtain
\beqn
R&=&\pi \theta  (\beta -1)\sum _{n=1}^\infty (-1)^n(\alpha ^{\frac{n}{2}}+
\alpha ^{-\frac{n}{2}})
\,\beta ^{-\frac{n}{2}}\,\left(-\frac{\ln \beta}{n}-\frac{2}{n^2}
+\frac{2}{n}\sum _{k=0}^{\infty}\left(\frac{1}{k+1}-\frac{1}{k+1+n}
\right)\right)
\nonumber \\
&&+\pi \theta (1-\beta )\sum _{n=1}^\infty (-1)^n(\alpha ^{\frac{n}{2}}+
\alpha ^{-\frac{n}{2}})
\,\beta ^{\frac{n}{2}}\,\left(\frac{\ln \beta}{n}-\frac{2}{n^2}
+\frac{2}{n}\sum _{k=0}^{\infty}\left(\frac{1}{k+1}-\frac{1}{k+1+n}
\right)\right)
\nonumber \\
&&+2\pi \theta (\beta -1)\sum _{n=-\infty}^\infty (-1)^n\alpha ^{\frac{n}{2}}
\,\sum _{k=0}^{\infty}\beta ^{-(k+1+\frac{|n|}{2})}\,\frac{1}{(k+1)(k+1+|n|)}
\nonumber\\
&&+2\pi \theta (1-\beta )\sum _{n=-\infty}^\infty (-1)^n\alpha ^{\frac{n}{2}}
\,\sum _{k=0}^{\infty}\beta ^{k+1+\frac{|n|}{2}}\,\frac{1}{(k+1)(k+1+|n|)}
\nonumber\\
&=&-\pi \ln \beta \, \sum _{n=1}^\infty \frac{(-1)^n}{n}(\alpha ^{\frac{n}{2}}+
\alpha ^{-\frac{n}{2}})
\,\left(\theta (\beta -1)\,\beta ^{-\frac{n}{2}}-\theta (1-\beta )
\beta ^{\frac{n}{2}}\right)
\nonumber\\
&&+2\pi \sum _{n=1}^\infty (-1)^n(\alpha ^{\frac{n}{2}}+
\alpha ^{-\frac{n}{2}})
\,\left(-\frac{1}{n^2}
+\frac{1}{n}\sum _{r=1}^{n}\frac{1}{r}
\right)\,\left(\theta (\beta -1)\,\beta ^{-\frac{n}{2}}+\theta (1-\beta )
\beta ^{\frac{n}{2}}\right)
\nonumber\\
&&+2\pi \theta (\beta -1)\left(\sum _{n=1}^\infty (-1)^n
\left(\alpha ^{\frac{n}{2}}+\alpha ^{-\frac{n}{2}}\right)
\,\sum _{m=1+\frac{n}{2}}^{\infty}\frac{\beta ^{-m}}{(m-\frac{n}{2})
(m+\frac{n}{2})}+\sum _{m=1}^{\infty}\frac{\beta ^{-m}}{m^2}\right)
\nonumber \\
&=&-\pi \ln \beta \, \sum _{n=1}^\infty \frac{(-1)^n}{n}(\alpha ^{\frac{n}{2}}+
\alpha ^{-\frac{n}{2}})
\,\left(\theta (\beta -1)\,\beta ^{-\frac{n}{2}}-\theta (1-\beta )
\beta ^{\frac{n}{2}}\right)
\nonumber\\
&&\pi \theta (\beta -1)\left(\left(\sum _{n=1}^\infty \frac{(-1)^n}{n}
\alpha ^{\frac{n}{2}}\beta ^{-\frac{n}{2}}\right)^2+
\left(\sum _{n=1}^\infty \frac{(-1)^n}{n}
\alpha ^{-\frac{n}{2}}\beta ^{-\frac{n}{2}}\right)^2\right)
\nonumber \\
&&+\pi \theta (1-\beta )\left(\left(\sum _{n=1}^\infty \frac{(-1)^n}{n}
\alpha ^{\frac{n}{2}}\beta ^{\frac{n}{2}}\right)^2+
\left(\sum _{n=1}^\infty \frac{(-1)^n}{n}
\alpha ^{-\frac{n}{2}}\beta ^{\frac{n}{2}}\right)^2\right)
\nonumber\\
&&+2\pi \theta (\beta -1)\sum _{n_1=1}^\infty \frac{(-1)^{n_1}}{n_1}
\alpha ^{\frac{n_1}{2}}\beta ^{-\frac{n_1}{2}}\,\sum _{n_2=1}^\infty
\frac{(-1)^{n_2}}{n_2}
\alpha ^{-\frac{n_2}{2}}\beta ^{-\frac{n_2}{2}}
\nonumber\\
&&+2\pi \theta (1-\beta )\sum _{n_1=1}^\infty \frac{(-1)^{n_1}}{n_1}
\alpha ^{\frac{n_1}{2}}\beta ^{\frac{n_1}{2}}\,\sum _{n_2=1}^\infty
\frac{(-1)^{n_2}}{n_2}
\alpha ^{-\frac{n_2}{2}}\beta ^{\frac{n_2}{2}}\,.
\eeqn
In obtaining two last contributions we passed to the new summation variables
$n_1=m+n/2$ and $n_2=m-n/2$. These transformations give a possibility to write the total result
for $R$ in the following simple form
\beqn
R&=&\pi \theta (\beta -1)\left(-\ln \beta \, \sum _{n=1}^\infty
\frac{(-1)^n}{n}
(\alpha ^{\frac{n}{2}}+
\alpha ^{-\frac{n}{2}})
\beta ^{-\frac{n}{2}}+\left(\sum _{n=1}^\infty \frac{(-1)^n}{n}
(\alpha ^{\frac{n}{2}}+
\alpha ^{-\frac{n}{2}})
\beta ^{-\frac{n}{2}}\right)^2\right)
\nonumber \\
&&+\pi \theta (1-\beta )\left(\ln \beta \, \sum _{n=1}^\infty \frac{(-1)^n}{n}
(\alpha ^{\frac{n}{2}}+
\alpha ^{-\frac{n}{2}})
\beta ^{\frac{n}{2}}+\left(\sum _{n=1}^\infty \frac{(-1)^n}{n}
(\alpha ^{\frac{n}{2}}+
\alpha ^{-\frac{n}{2}})
\beta ^{\frac{n}{2}}\right)^2\right)
\nonumber \\
&=&\pi \theta (\beta -1)\,\left(\sum _{n=1}^\infty \frac{(-1)^n}{n}
(\alpha ^{\frac{n}{2}}+
\alpha ^{-\frac{n}{2}})
\beta ^{-\frac{n}{2}}\right)\left(\sum _{n=1}^\infty \frac{(-1)^n}{n}
(\alpha ^{\frac{n}{2}}+
\alpha ^{-\frac{n}{2}})
\beta ^{-\frac{n}{2}}-\ln \beta\right)
\nonumber \\
&&+\pi \theta (1-\beta )\,\left(\sum _{n=1}^\infty \frac{(-1)^n}{n}
(\alpha ^{\frac{n}{2}}+
\alpha ^{-\frac{n}{2}})
\beta ^{\frac{n}{2}}\right)\left(\sum _{n=1}^\infty \frac{(-1)^n}{n}
(\alpha ^{\frac{n}{2}}+
\alpha ^{-\frac{n}{2}})
\beta ^{\frac{n}{2}}+\ln \beta\right)
\nonumber \\
&=&\pi \theta (\beta -1)\,\ln \left((1+\sqrt{\alpha}/\sqrt{\beta})
(1+1/\sqrt{\alpha \beta })\right)
\left(\ln \left((1+\sqrt{\alpha}/\sqrt{\beta})
(1+1/\sqrt{\alpha \beta })\right)+\ln \beta \right)
\nonumber \\
&&+\pi \theta (1-\beta )\,\ln \left((1+\sqrt{\beta}/\sqrt{\alpha})
(1+\sqrt{\alpha \beta })\right)
\left(\ln \left((1+\sqrt{\beta}/\sqrt{\alpha})
(1+\sqrt{\beta \alpha })\right)-\ln \beta \right).
\eeqn
Using finally the above expression for $M_{n \neq 0}$ we obtain
\beqn
R=\pi \ln \frac{|k_1+k_2|^2|q_2|^2}{|k_2|^2|q_1|^2}\,
\ln \frac{|k_1+k_2|^2|q_2|^2}{|k_1|^2|q_3|^2}\,.
\eeqn
Combination of the one and two loop results leads to: 
\beqn
\frac{1}{\pi}\Im _{s_2}\,M_{2\rightarrow 4} =  \frac{a}{4\pi  }\,
s_2^{\omega (t_2)}\sum _{n=-\infty}^\infty (-1)^n Reg _{s_2}
\int _{-\infty}^\infty \frac{d \nu}{\nu ^2+\frac{n^2}{4}}\,
\left(\frac{q_3^*k_1^*}{k_2^*q_1^*}\right)^{i\nu -\frac{n}{2}}
\left(\frac{q_3k_1}{k_2q_1}\right)^{i\nu +\frac{n}{2}}
s_2^{\omega (\nu ,n)}\,
\nonumber \\
=  \frac{a}{2  }\,
s_2^{\omega (t_2)}\,\left(\ln \frac{|k_1|^2|k_2|^2}{|k_1+k_2|^2\mu ^2}
-\frac{1}{\epsilon}-\frac{a}{2}\ln s_2 \,\ln
\frac{|k_1+k_2|^2|q_2|^2}{|k_2|^2|q_1|^2}\,
\ln \frac{|k_1+k_2|^2|q_2|^2}{|k_1|^2|q_3|^2} + {\cal O}(a^2)\right).
\eeqn
Indeed, the second term of the expansion in $a$ coincides
with the result (63), obtained by an independent calculation.

In an analogous way the imaginary part in $s$ can be written in the form
\beq
\frac{1}{\pi}\Im _{s}\,M_{2\rightarrow 4}=\frac{a}{4\pi  }\,
s_2^{\omega (t_2)}\sum _{n=-\infty}^\infty (-1)^n Reg _{s}
\int _{-\infty}^\infty \frac{d \nu}{\nu ^2+\frac{n^2}{4}}\,
\left(\frac{q_3^*k_1^*}{k_2^*q_1^*}\right)^{i\nu -\frac{n}{2}}
\left(\frac{q_3k_1}{k_2q_1}\right)^{i\nu +\frac{n}{2}}
s_2^{\omega (\nu ,n)}\,,
\eeq
where
\beq
Reg _{s}\int _{-\infty}^\infty \frac{d \nu}{\nu ^2}\,
\left|\frac{q_3k_1}{k_2q_1}\right|^{2i\nu}=2\pi \left(-\frac{1}{\epsilon}
+\ln \frac{|q_2|^2}{\mu ^2} +
\ln \left|\frac{q_3q_1}{k_1k_2}\right|
- \left|\ln \left|\frac{q_3k_1}{q_1k_2}\right|\right|
\right)\,.
\eeq
It corresponds to the following expansion in $a$
\begin{align}
\frac{1}{\pi}\Im _{s}\,M_{2\rightarrow 4}=\hspace{11cm}\nonumber\\
\frac{a}{2  }\,
s_2^{\omega (t_2)}\,\left(\ln \frac{|q_1|^2|q_3|^2}{|k_1+k_2|^2\mu ^2}
-\frac{1}{\epsilon}-\frac{a}{2}\ln s_2 \,\ln
\frac{|k_1+k_2|^2|q_2|^2}{|k_2|^2|q_1|^2}\,
\ln \frac{|k_1+k_2|^2|q_2|^2}{|k_1|^2|q_3|^2} + {\cal O}(a^2) \right).
\end{align}

For the imaginary part of the amplitude $M_{3\rightarrow 3}$ in the
variable $t'_2$ we obtain the similar result
\beq
\frac{1}{\pi}\Im _{t'_2}\,M_{3\rightarrow 3}=\frac{a}{4\pi  }\,
t_2^{\prime \,\omega (t_2)}\sum _{n=-\infty}^\infty (-1)^n Reg _{t'_2}
\int _{-\infty}^\infty \frac{d \nu}{\nu ^2+\frac{n^2}{4}}\,
\left(\frac{q_3^*q_1^*}{k_2^*k_1^*}\right)^{i\nu -\frac{n}{2}}
\left(\frac{q_3q_1}{k_2k_1}\right)^{i\nu +\frac{n}{2}}
t_2^{\prime \omega (\nu ,n)}\,,
\eeq
where in this case the regularized integral over $\nu$ for $n=0$ and $a=0$
does not contain any $1/\epsilon$ divergence
\beq
Reg _{t'_2}\int _{-\infty}^\infty \frac{d \nu}{\nu ^2}\,
\left|\frac{q_3q_1}{k_2k_1}\right|^{2i\nu}=2\pi \left(  
\ln \left|\frac{q_3q_1}{k_1k_2}\right| - \left|\ln \left|\frac{q_3q_1}{k_1k_2}\right|\right|
\right)
\,.
\eeq
It gives the following $a$-expansion of $\Im _{t'_2}\,M_{3\rightarrow 3}$
\begin{align}
\frac{1}{\pi}\Im _{t'_2}\,M_{3\rightarrow 3}=\hspace{12.5cm} \nonumber\\
\frac{a}{2  }\,
t_2^{\prime \,\omega (t_2)}\,\left(\ln
\frac{|q_1|^2|q_3|^2}{|q_1+q_3-q_2|^2|q_2| ^2}
-\frac{a}{2}\ln t'_2 \,\ln
\frac{|q_1+q_3-q^2|^2|q_2|^2}{|k_2|^2|k_1|^2}\,
\ln \frac{|q_1+q_3-q_2|^2|q_2|^2}{|q_1|^2|q_3|^2} + {\cal O}(a^2) \right).
\end{align}
Analogously we find the the imaginary part of the amplitude
$M_{3\rightarrow 3}$ in the variable $s$
\beq
\frac{1}{\pi}\Im _{s}\,M_{3\rightarrow 3}=\frac{a}{4\pi  }\,
t_2^{\prime \omega (t_2)}\sum _{n=-\infty}^\infty (-1)^n Reg _{s}
\int _{-\infty}^\infty \frac{d \nu}{\nu ^2+\frac{n^2}{4}}\,
\left(\frac{q_3^*q_1^*}{k_2^*k_1^*}\right)^{i\nu -\frac{n}{2}}
\left(\frac{q_3q_1}{k_2k_1}\right)^{i\nu +\frac{n}{2}}
t_2^{\prime \omega (\nu ,n)}\,,
\eeq
where
\beq
Reg _{s}\int _{-\infty}^\infty \frac{d \nu}{\nu ^2}\,
\left|\frac{q_3q_1}{k_2k_1}\right|^{2i\nu}=- 2\pi 
\left( - \ln \left|\frac{q_3q_1}{k_1k_2}\right| -\left|\ln \left|\frac{q_3q_1}{k_1k_2}\right|
\right| \right)
\,.
\eeq
The expansion in $a$ beyond the one loop approximation coincides with that 
of $\Im _{t'_2}\,M_{3\rightarrow 3}$:
\begin{align}
\frac{1}{\pi}\Im _{s}\,M_{3\rightarrow 3}=\hspace{12.5cm} \nonumber\\
\frac{a}{2  }\,
t_2^{\prime \,\omega (t_2)}\,\left(\ln
\frac{|k_1|^2|k_2|^2}{|q_1+q_3-q_2|^2|q_2| ^2}
-\frac{a}{2}\ln t'_2 \,\ln
\frac{|q_1+q_3-q^2|^2|q_2|^2}{|k_2|^2|k_1|^2}\,
\ln \frac{|q_1+q_3-q_2|^2|q_2|^2}{|q_1|^2|q_3|^2} + {\cal O}(a^2) \right).
\end{align}

\end{document}